\documentclass[aps,twocolumn,showpacs,pra,floatfix]{revtex4}

\usepackage{amsfonts,amsmath,amssymb,mathrsfs,amsthm}
\usepackage{dsfont}
\usepackage{graphicx}

\usepackage{setspace}
\usepackage{url}
\usepackage{bm}

\usepackage{color}

\newtheorem{thm}{Theorem}
\newtheorem{thm1}{Theorem}
\newtheorem{thm2}{Theorem}
\newtheorem{thm3}{Theorem}
\newtheorem{thm4}{Theorem}
\newtheorem{conjec}[thm1]{Conjecture}
\newtheorem{ex}[thm2]{Example}
\newtheorem{defn}[thm3]{Definition}
\newtheorem{rem}[thm4]{Remark}
\newcommand{\bd}[1]{\emph{\textbf{#1}}}
\newcommand{\captionC}[1]{\caption{\emph{#1}}}

\begin{document}

\title{Quasipinning and its relevance for $N$-Fermion quantum states}

\author{Christian Schilling}
\email{christian.schilling@physics.ox.ac.uk}
\affiliation{Clarendon Laboratory, University of Oxford, Parks Road, Oxford OX1 3PU, United Kingdom}

%

\date{\today}

\begin{abstract}
Fermionic natural occupation numbers (NON) do not only obey Pauli's famous exclusion principle but are even further restricted to a polytope by the generalized Pauli constraints, conditions which follow from the fermionic exchange statistics. Whenever given NON are pinned to the polytope's boundary the corresponding $N$-fermion quantum state $|\Psi_N\rangle$ simplifies due to a selection rule. We show analytically and numerically for the most relevant settings that this rule is stable for NON close to the boundary, if the NON are non-degenerate. In case of degeneracy a modified selection rule is conjectured and its validity is supported.
As a consequence the recently found effect of quasipinning is physically relevant in the sense that its occurrence allows to approximately reconstruct $|\Psi_N\rangle$, its entanglement properties and correlations from 1-particle information. Our finding also provides the basis for a generalized Hartree-Fock method by a variational ansatz determined by the selection rule.
\end{abstract}

\pacs{03.67.-a, 05.30.Fk, 31.15.B-}

\maketitle

\section{Introduction}\label{sec:intro}
The rigorous treatment of $N$-particle quantum systems is typically impossible from an analytic viewpoint and at least very challenging from a numerical one. In particular for macroscopic particle numbers this task is absolutely hopeless. As a consequence effective reduced descriptions have been developed. Prime examples are mean field approximations for identical particles like the Hartree-Fock approximation describing the physical behavior of a single particle in the self-consistent background field generated by all the other particles. Even more rudimentary is the starting point for the description of many effects in condensed matter physics. There, the interaction between the electrons is switched off and the corresponding time independent $N$-electron Schr\"odinger equation reduces to a $1$-electron Schr\"odinger equation, describing a single electron in the crystal field of the nuclei. The interaction between the electrons is then reintroduced by perturbation theoretical means (see e.g.~\cite{FW03}). Besides the essential numerical simplifications of those $1$-electron approximations there is also an important related conceptual simplification. Going from the complicated $N$-particle to the more elementary $1$-particle picture can allow one to gain some insight into the physics of the problem at hand. Indeed, we can easily imagine single particles occupying energy shells as e.g.~in atoms or lattice sites and their time evolution is given by the hopping of the particles between energy shells or lattice sites. This is in contrast to the $N$-particle picture since the collective many-particle behavior is often beyond any imagination.

Since physical effects are typically based on $2$-particle interactions it is \emph{a priori} surprising that $1$-particle methods as e.g.~the Hartree-Fock method can provide at least for some physical systems of interacting electrons reasonable results. As an example the ground state energies of atoms are obtained by the Hartree-Fock ansatz with an error of at most a few percent which even reduces for atoms with larger atomic numbers \cite{Bach1992}. Due to the success of the $1$-particle picture kinematical constraints on occupation numbers are fundamentally important. This is the reason why Pauli's famous exclusion principle \cite{Pauli1925} has often a strong impact on the behavior and the properties of electronic systems. By restricting the occupancies of quantum states some electronic transitions in atoms are not possible anymore. In the same way the behavior of solid state  materials at low temperatures is determined by the electrons around the Fermi level. Mathematically, Pauli's exclusion principle can be stated as
\begin{equation}\label{PauliExn}
0\leq n_{\varphi}\leq 1\,,
\end{equation}
where $n_{\varphi}$ is the particle number expectation value for some $1$-particle state $\varphi$.
In 1926 Pauli's exclusion principle was identified by Dirac and Heisenberg \cite{Dirac1926, Heis1926} as a consequence of the antisymmetry of the corresponding $N$-fermion quantum state under particle exchange. Only in 1972 has it been shown, by Borland and Dennis \cite{Borl1972}, that for a system of $3$ fermions and a $6$-dimensional $1$-particle Hilbert space the antisymmetry of the $3$-fermion quantum state leads to further restrictions on natural occupation numbers. Only recently it was shown \cite{Kly2,Kly3,Altun} that for settings of arbitrary fermion number $N$ and arbitrary dimension $d$ of the $1$-particle Hilbert space one has also extra constraints on natural occupation numbers. Klyachko and Altunbulak \cite{Kly3,Altun} provide an algorithm which allows for each fixed $N$ and $d$ to calculate all restrictions. These so-called generalized Pauli constraints are significantly stronger than Pauli's exclusion principle and give rise to a polytope of mathematically possible occupation numbers.

The central question for physical applications of this new structure is whether concrete fermionic states as e.g.~ground states are pinned to the boundary of the polytope, i.e.~their occupation numbers saturate one of these linear inequalities. It has been claimed
in \cite{Kly1} that pinning as an effect in the $1$-particle picture allows one to reconstruct the structure of the corresponding $N$-fermion quantum state $|\Psi_N\rangle$. This is a generalization of the well-known result that a set of occupation numbers all equal to $1$ or $0$ can just arise from a single Slater determinant $|\Psi_N\rangle = |1,2,\ldots,N\rangle$. The latter statement is also stable in the sense that occupation numbers all \emph{approximately} equal to $1$ or $0$ imply $|\Psi_N\rangle \approx |1,2,\ldots,N\rangle$ \cite{Bach1992}.
Although pinning seems to be a quite spectacular effect, analytic evidence was provided recently in \cite{CS2013} that fermionic ground states may exhibit quasipinning rather than pinning. There, it has been shown for a particular physical system that the occupation numbers are very close, but not exactly on the polytope boundary. This new effect is only physically relevant when the structural implications of pinning are stable in the vicinity of the polytope boundary and therefore also apply (approximately) for quasipinning. To explore and eventually verify this kind of stability is our goal. In that sense we also emphasize the relevance of the quite recent development in quantum physics and quantum chemistry \cite{Kly1,BenavLiQuasi,Mazz14}, the investigation of possible (quasi)pinning of atomic states.

Our work is arranged as follows. In Sec.~\ref{sec:notation} we introduce some useful concepts for fermionic quantum systems, in particular an exact self-consistent expansion of $N$-fermion quantum states in Slater determinants built up from its own natural orbitals as well as a geometric picture for the description of fermionic occupation numbers. Sec.~\ref{sec:gPC} briefly reviews the concept of generalized Pauli constraints. In Sec.~\ref{sec:physrev} we explain that pinning is physically relevant in the sense that it implies strong structural implications for the corresponding $N$-fermion quantum state. In the main part, Sec.~\ref{sec:quasipinning}, we verify that quasipinning implies approximately these structural simplifications of pinning and eventually conclude that quasipinning is also physically relevant. This is first shown analytically for the so-called Borland-Dennis setting of $3$ fermions and a $6$-dimensional $1$-particle Hilbert space and then by a numerical study extended to larger settings. There we analyze randomly sampled quantum states and verify possible structural implications for those states which exhibit quasipinning.

\section{Concepts for fermionic quantum systems}\label{sec:notation}
In order to keep our presentation self-contained we introduce some notation, the concept of natural orbitals and natural occupation numbers as well as a self-consistent expansion
of $N$-fermion quantum states. This expansion in Slater determinants built up from its own natural orbitals allows to study structural aspects of fermionic quantum states more elegantly and suggests a geometric picture for the description of natural occupation numbers.

We begin by considering $N$ identical fermions. Their quantum states, assumed onwards to be pure, are given by elements in the $N$-fermion Hilbert space
\begin{equation}
|\Psi_N\rangle \in \wedge^N[\mathcal{H}_1^{(d)}] \equiv \mathcal{H}_N^{(f)}
\end{equation}
of antisymmetric $N$-particle states, where we assume the $1$-particle Hilbert space $\mathcal{H}_1^{(d)}$ to be finite, $d$-dimensional. The concrete spatial form of the states in $\mathcal{H}_1^{(d)}$ as well as the Hamiltonian describing the physics of that system is not relevant for the following considerations. For a given $N$-fermion state $|\Psi_N\rangle$
the $1$-particle reduced density operator ($1$-RDO) $\rho_1$ is defined by tracing out $N-1$ fermions,
\begin{equation}\label{rho1}
\rho_1 \equiv N\,\mbox{Tr}_{N-1}[|\Psi_N\rangle\langle\Psi_N|]\,.
\end{equation}
The partial trace in Eq.~(\ref{rho1}) is well-defined, which follows directly from the natural embedding $\wedge^N[\mathcal{H}_1^{(d)}] \leq \left(\mathcal{H}_1^{(d)}\right)^{\otimes^N}\equiv \mathcal{H}_N$ of $\wedge^N[\mathcal{H}_1^{(d)}]$ into the $N$-particle Hilbert space $\mathcal{H}_N$
(without any exchange-symmetry). Moreover, due to the specific exchange symmetry it does not matter which $N-1$ fermions we trace out. Diagonalizing $\rho_1$,
\begin{equation}\label{rho1diag}
\rho_1 = \sum_{k=1}^d\,\lambda_k\,|k\rangle \langle k|\,,
\end{equation}
gives rise to the \emph{natural occupation numbers} (NON) $\lambda_k$ and \emph{natural orbitals} (NO) $|k\rangle$, the corresponding eigenstates. This terminology also motivates the normalization $\mbox{Tr}_1[\rho_1]=\lambda_1+\ldots+\lambda_d =N$ allowing us to interpret the eigenvalues of $\rho_1$ as occupation numbers. Consequently, Pauli's famous exclusion principle can be formulated as
\begin{equation}\label{PauliEx}
0\leq \lambda_k\leq 1\qquad,\,\forall k=1,2,\ldots,d\,.
\end{equation}
Moreover, in Eq.~(\ref{rho1diag}) we order the NON decreasingly, $\lambda_1\geq \lambda_2\geq\ldots\geq \lambda_d\geq 0$.
The NO give rise to an orthonormal basis $\mathcal{B}_1\equiv \{|k\rangle\}_{k=1}^d$ for the $1$-particle Hilbert space $\mathcal{H}_1^{(d)}$.
This basis is unique (up to global phases) as long as the NON are non-degenerate, which we typically assume, if not stated differently. $\mathcal{B}_1$
induces an orthonormal basis $\mathcal{B}_N$ for $\mathcal{H}_N^{(f)}$, the family of Slater determinants,
\begin{equation}
|\emph{\textbf{i}}\,\rangle \equiv \mathcal{A}_N[|i_1\rangle\otimes \ldots \otimes |i_N\rangle]\,,
\end{equation}
where $\emph{\textbf{i}} \equiv(i_1,\ldots,i_N)$, $1\leq i_1<i_2< \ldots <i_N\leq d$ and $\mathcal{A}_N$ is the antisymmetrizing operator on $\mathcal{H}_N$.
Just for ease of notation we skip the explicit dependence of the elements in $\mathcal{B}_1$ and $\mathcal{B}_N$ on $|\Psi_N\rangle$\,.
Since $\mathcal{B}_N$ is a basis for $\mathcal{H}_N^{(f)}$ we can expand every state in $\mathcal{H}_N^{(f)}$ uniquely w.r.t.~$\mathcal{B}_N$, in particular also $|\Psi_N\rangle$,
\begin{equation}\label{Psic}
|\Psi_N\rangle = \sum_{\bd{i}}\, c_{\bd{i}}\,|\bd{i}\,\rangle\,,
\end{equation}
where we recall that $|\bd{i}\,\rangle$ depends on $|\Psi_N\rangle$.
Notice that the self-consistency of this expansion imposes quite strong restrictions on the set of expansion coefficients $c_{\bd{i}}$.
They reflect the fact that the corresponding $1$-RDO (\ref{rho1}) is diagonal w.r.t.~the NO $|k\rangle$. In addition the occupation number of $|k\rangle$
is $\lambda_k$, the $k$-th largest NON. As a consequence of the self-consistent notation, we can express the NON as an elementary function
of the expansion coefficients $c_{\bd{i}}$. By denoting the particle number operator of NO $|k\rangle$ by $\hat{n}_k$ we find
\begin{eqnarray}\label{NONc}
\lambda_k &=& \langle \Psi_N|\hat{n}_k|\Psi_N\rangle\\
&=&\sum_{\bd{i},\bd{j}}\,c_{\bd{i}}^\ast\,c_{\bd{j}}\,\langle \bd{i}\,|\hat{n}_k|\bd{j}\,\rangle
= \sum_{\bd{i}}\,|c_{\bd{i}}|^2\,\langle \bd{i}\,|\hat{n}_k|\bd{i}\,\rangle
=\sum_{\bd{i}, k\in \bd{i}}\,|c_\bd{i}|^2\,.  \nonumber
\end{eqnarray}

Eqs.~(\ref{Psic}), (\ref{NONc}) allow us to introduce a geometric picture for the description of NON, $\vec{\lambda} \equiv (\lambda_1,\ldots,\lambda_d)
\in \mathbb{R}^d$. The Pauli exclusion principle Eq.~(\ref{PauliEx}) restricts the space of possible $\vec{\lambda}$ to the Pauli hypercube
\begin{equation}\label{Paulicube}
\mathcal{C}\equiv [0,1]^d \subset\mathbb{R}^d\,.
\end{equation}
Of course, the ordering of the NON leads to a further restriction and the normalization of the NON defines a cut through $\mathcal{C}$.
We denote the $2^d$ vertices of $\mathcal{C}$ by $\vec{v}_{\bd{l}}$, where $\bd{l}\subset\{1,2,\ldots,d\}$ and
\begin{equation}\label{vertices1}
\forall j=1,2,\ldots,d\,:\qquad(\vec{v}_{\bd{l}})_j =\begin{cases}
  1,  &  j \in \bd{l}\\
  0, & j \not \in \bd{l}
\end{cases}\,.
\end{equation}
However, according to Eq.~(\ref{NONc}), the vector $\vec{\lambda}$ of NON is spanned only by those $\bd{i}\subset \{1,\ldots,d\}$ which have length $N$.
We denote that set by $\mathcal{I}_N$ and can rewrite Eq.~(\ref{NONc}) as
\begin{equation}\label{NONgeom}
\vec{\lambda} = \sum_{\bd{i} \in \mathcal{I}_N}\, |c_{\bd{i}}|^2\, \vec{v}_{\bd{i}}\,.
\end{equation}
This means that $\vec{\lambda}$ is the ``center of mass'' for masses $\{|c_{\bd{i}}|^2\,|\,\bd{i}\in \mathcal{I}_N\}$ sitting at the $\binom{d}{N}$ vertices $\vec{v}_{\bd{i}}$, $\bd{i}\in \mathcal{I}_N$. Moreover, we can identify those vertices with the $N$-particle Slater determinants. By recalling Eq.~(\ref{NONc}), Eq.~(\ref{vertices1}) for vertices $\vec{v}_{\bd{i}}$ with $\bd{i} \in \mathcal{I}_N$ becomes
\begin{equation}
(\vec{v}_{\bd{i}})_j = \langle \bd{i}\,|\hat{n}_j|\bd{i}\,\rangle\,.
\end{equation}
This means that $\vec{v}_{\bd{i}}$ is the list of occupation numbers of the Slater determinant $|\bd{i}\,\rangle$ w.r.t.~the NO $\{|j\rangle\}_{j=1}^d$ of the given state $|\Psi_N\rangle$. This geometric picture is illustrated and intensively used (in an even simpler form) in Sec.~\ref{sec:BDsetting}, by studying the setting $\wedge^3[\mathcal{H}_1^{(6)}]$.

For a given orthonormal basis $\mathcal{B}_1=\{|i\rangle\}_{i=1}^d$ for the $1$-particle Hilbert space $\mathcal{H}_1^{(d)}$ we  introduce the corresponding fermionic creation and annihilation operators $a_i^\dagger, a_i$, which create and annihilate a fermion in the state $|i\rangle$. They fulfill the anticommutation relations
\begin{equation}
\{a_j,a_k\} = \{a_j^\dagger,a_k^\dagger\} = 0\,,\qquad \{a_j,a_k^\dagger\} = \delta_{j k}\,.
\end{equation}
Using this second quantization we can also express the $1$-RDO as
\begin{equation}
\langle j|\rho_1|k\rangle = \langle \Psi_N|a_j^\dagger a_k |\Psi_N\rangle\qquad,\,\forall j,k=1,\ldots,d\,.
\end{equation}
In particular, if we choose as $1$-particle basis the NO of a given quantum state $|\Psi_N\rangle$ we find for its NON
\begin{equation}\label{NONexpect}
\lambda_j = \langle \Psi_N| a_j^\dagger a_j|\Psi_N\rangle\,.
\end{equation}
Since all particle number operators $\hat{n}_j= a_j^\dagger a_j$ commute with each other any function $F(\vec{\lambda})$ can be expressed
as expectation value
\begin{equation}\label{Functexpect}
F(\vec{\lambda}) = \langle \Psi_N| \hat{F}|\Psi_N\rangle
\end{equation}
with
\begin{equation}\label{Fop}
\hat F \equiv F(a_1^\dagger a_1,\ldots, a_d^\dagger a_d)\,.
\end{equation}
In the following, we frequently use this concept of assigning an operator $\hat{F}$ to any function
$F(\vec{\lambda})$. We call $\hat{F}$ the operator of $F$ w.r.t.~the NO of the  given
state $|\Psi_N\rangle$ or just the operator of $F$. Typically, we will also skip the index ``$\Psi_N$'' for
operators $a_j^\dagger, a_k, \hat F,\ldots$ expressing their dependence on the underlying quantum state
$|\Psi_N\rangle$.

\section{Generalized Pauli constraints and pinning}\label{sec:gPC}
\begin{figure}[h]
\centering
\includegraphics[scale=0.6]{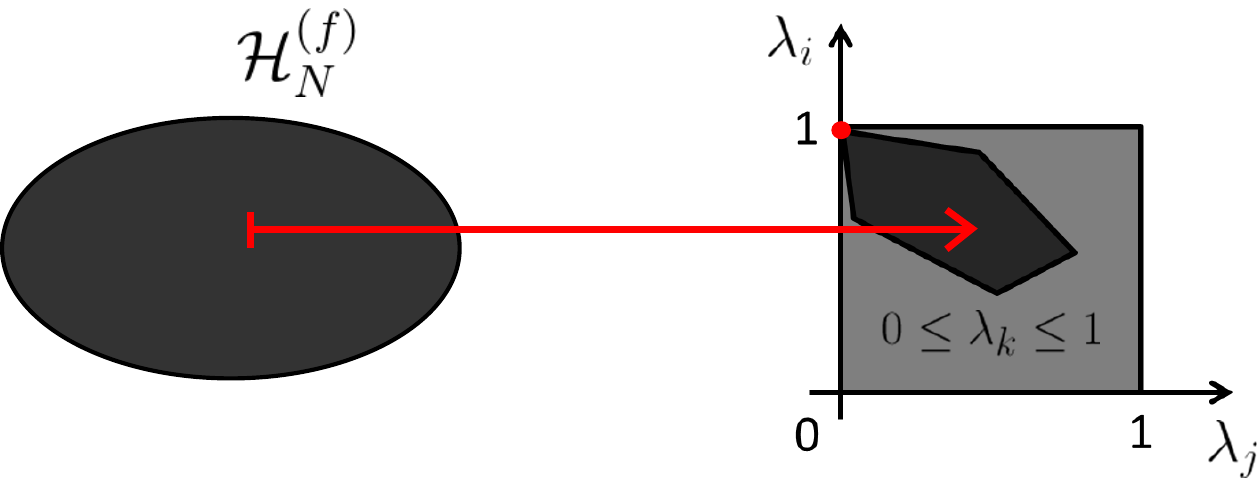}
\caption{The family of antisymmetric $N-$particle states depicted as an ellipse maps to the family of possible natural occupation numbers $\vec{\lambda}$ (dark-gray), which turns out to be a proper subset of the Pauli hyper cube (light-gray). The ``Hartree-Fock point'' is shown as a red dot.}
\label{fig:fqmp}
\end{figure}

The Pauli exclusion principle (\ref{PauliEx}) is an elementary consequence of the antisymmetry of $N$-fermion quantum states under particle exchange.
Since the antisymmetry is a much stronger mathematical condition than Pauli's exclusion principle alone, it is not surprising that it leads to further restrictions on NON strengthening the exclusion principle. This is illustrated in Fig.~\ref{fig:fqmp}.
As explained in Sec.~\ref{sec:notation} each $N$-fermion quantum state $|\Psi_N\rangle$ can be mapped to its vector $\vec{\lambda}=(\lambda_i)_{i=1}^d$ of NON. These NON do not only lie inside of the light-gray Pauli hypercube (\ref{Paulicube}) but are further restricted to the dark-gray polytope (see Fig.~\ref{fig:fqmp}).
This polytope $\mathcal{P}_{N,d}$ is determined by a finite family of so called \emph{generalized Pauli constraints} (gPC), linear conditions
\begin{equation}\label{gPC}
D_i^{(N,d)}(\vec{\lambda}) = \kappa_i^{(0)}  + \vec{\kappa}_i \cdot \vec{\lambda} \geq 0\,,
\end{equation}
with $\vec{\kappa}_i \equiv (\kappa_i^{(j)})_{j=1}^d$, $\kappa_i^{(j)} \in \mathbb{Z}$, $j=0,1,\ldots,d$ and $i=1,2,\ldots,r^{(N,d)}$.
The number $r^{(N,d)}$ of such constraints (\ref{gPC}) increases drastically with the dimension $d$ of the $1$-particle Hilbert space.
In contrast to Pauli's exclusion principle, the family of gPC depends on the number $N$ of fermions
and the dimension $d$ of the $1$-particle Hilbert space. In a ground-breaking work by Klyachko and Altunbulak \cite{Kly3,Altun} an algorithm was found,
which allows one to calculate for each fixed $N$ and $d$ those constraints.

As an example we present the gPC for the setting $\wedge^3[\mathcal{H}_1^{(6)}]$, which were already found in 1972 by Borland and Dennis \cite{Borl1972}
\begin{eqnarray}
&&\lambda_1+\lambda_6 = \lambda_2+\lambda_5 = \lambda_3+\lambda_4 = 1\,, \label{d=6a} \\
&&D^{(3,6)}(\vec{\lambda}) \equiv 2-(\lambda_1 +\lambda_2+\lambda_4) \geq 0 \label{d=6b} \,.
\end{eqnarray}
We remind the reader that the NON are always ordered decreasingly.
Notice that the inequality $D^{(3,6)}(\vec{\lambda}) \geq 0$ is stronger than Pauli's exclusion principle, which just states
that $2-(\lambda_1 +\lambda_2)\geq 0$. That some constraints take the form of equalities (instead of inequalities) as those in (\ref{d=6a}) is specific
and happens only for the quite ``small'' settings $\wedge^3[\mathcal{H}_1^{(6)}]$, $\wedge^2[\mathcal{H}_1^{(d)}]$ and
$\wedge^{d-2}[\mathcal{H}_1^{(d)}]$. E.g.~for the settings $\wedge^2[\mathcal{H}_1^{(d)}]$ the only restriction of NON is given by the condition that every non-zero NON is evenly degenerate, $\lambda_{2j-1}=\lambda_{2j}$, $j=1,2,\ldots$. Although gPC taking the form of equalities are much more restrictive than those taking the form of inequalities we are, at least from a physical viewpoint, more interested in the latter ones. This is because their potential (quasi)saturation defines a characteristic feature which may emerge from a physical mechanism, as e.g.~the energy minimization.

Due to the particle-hole duality (see e.g.~\cite{RusEquiv}) which is based on a natural isomorphism $\wedge^N[\mathcal{H}_1^{(d)}] \cong\wedge^{d-N}[\mathcal{H}_1^{(d)}]$ the polytope $\mathcal{P}_{d-N,d}$ follows from $\mathcal{P}_{N,d}$ by replacing $\lambda_i \mapsto 1-\lambda_{d-i+1}$, $i=1,2,\ldots,d$. Consequently, the Borland-Dennis setting is for our purpose the first non-trivial setting. In the following if we consider an arbitrary setting $\wedge^N[\mathcal{H}_1^{(d)}]$ with polytope $\mathcal{P}_{N,d}$ we assume w.l.o.g.~$d\geq 2N$
and skip the superscript $(N,d)$ of the gPC $D_j^{(N,d)}$ as well as its labeling index $j$.

It is important to notice that the polytope $\mathcal{P}_{N,d}$ emerges just from kinematics (antisymmetry of $|\Psi_N\rangle$)
and is independent of the underlying Hamiltonian which is responsible for the dynamics of the concrete physical system. To explore the
physical implications of this new mathematical structure revealed by Klyachko and Altunbulak the first task is to understand where vectors
$\vec{\lambda}$ of occupation numbers of relevant fermionic quantum states do lie. If we
consider e.g.~ground states of non-interacting fermions confined by some
external potential the position of the corresponding $\vec{\lambda}$-vector is obvious. The ground state is given by a single Slater determinant $|1,2,\ldots,N\rangle$, where each of the lowest $N$ energy states of the external trap
is occupied by one fermion. Consequently, the corresponding NON are given by $\vec{\lambda}=(1,\ldots,1,0,\ldots)$. This point is shown in Fig.~\ref{fig:fqmp} as red dot. If we turn on some interaction with coupling strength
$\delta$ the NON $\vec{\lambda}(\delta)$ will move away from the so-called ``Hartree-Fock point'' $(1,\ldots,1,0,\ldots)$.
The central question is then whether $\vec{\lambda}(\delta)$ moves towards the middle of the polytope or whether it still lies on the boundary
of the polytope. In the latter case we say that the NON are \emph{pinned to the boundary} of the polytope, an effect first time mentioned by Klyachko in \cite{Kly1}. Mathematically, pinning of NON $\vec{\lambda}_0$ is defined by the saturation of a gPC $D(\vec{\lambda})\geq 0$,
\begin{equation}
D(\vec{\lambda}_0) = 0\,.
\end{equation}
Geometrically, this means that $\vec{\lambda}_0$ lies on the corresponding facet $F_D$ of the polytope $P_{N,d}$ defined by
\begin{equation}\label{facet}
F_D = \{\vec{\lambda}\in \mathcal{P}_{N,d}\,|\,D(\vec{\lambda})=0\}\,.
\end{equation}
For applications below, it is essential to distinguish between the saturation of gPC $D(\vec{\lambda})\geq 0$ and that of the ordering
constraints $\lambda_i -\lambda_{i+1} \geq 0$. Consequently, we also need to distinguish between the facets $F_D$ of $P_{N,d}$ corresponding to the
saturation of $D(\vec{\lambda})\geq 0$  and the remaining facets defined by $\lambda_i = \lambda_{i+1}$. Since it will turn out that only the saturation of gPC has some physical meaning we restrict the concept of pinning just to gPC. In particular,
if we say that the NON $\vec{\lambda}$ are lying on the boundary of the polytope $P_{N,d}$ we have in mind that it lies on a facet $F_D$ corresponding to some gPC $D(\vec{\lambda})\geq 0$.

\section{Potential physical relevance of pinning}\label{sec:physrev}
In this section we explain the potential physical relevance of pinning. Two quite elementary reasons were already mentioned in \cite{Kly1,CSQMath12}:
\begin{enumerate}
\item For a pinned ground state some gPC may be active for the energy minimization in the sense that any further minimization would violate it.
      In that sense this gPC would have a strong impact on the ground state energy.
\item For a quantum system initially prepared in a pinned state $|\Psi_N\rangle$ its time evolution $|\Psi_N(t)\rangle$ from the viewpoint of the $1$-particle picture is limited. Since the corresponding vector $\vec{\lambda}(t)$ of NON can never leave the polytope it cannot evolve in an arbitrary direction.
\end{enumerate}

Besides those two reasons for a possible relevance of pinning for concrete physical systems a central question is whether pinning as effect in the $1$-particle picture provides any information about the corresponding $N$-fermion quantum state $|\Psi_N\rangle$. That this is not absurd is indicated by a well-known result (see e.g.~\cite{Bach1992}). Any $N$-fermion quantum state $|\Psi_N\rangle$ with NON
\begin{equation}\label{NONHF}
\vec{\lambda} = (\underbrace{1,\ldots,1}_N,0,\ldots) \equiv \vec{\lambda}_{HF}
\end{equation}
can be written as one single Slater determinant, $|\Psi_N\rangle = |1,2,\ldots,N\rangle$.
Even more important is the stability of this structural implication. Indeed, it can easily be shown (see e.g.~\cite{Bach1992,SuplMat})
that $|\Psi_N\rangle$ with NON $\vec{\lambda} \approx \vec{\lambda}_{HF}$ can approximately be written
as one single Slater determinants, $|\Psi_N\rangle \approx |1,2,\ldots,N\rangle$. This stability is the reason why the Hartree-Fock method is a meaningful approximation.

Now we explain that this structural implication can be generalized to pinning of NON to an \emph{arbitrary} point on the polytope boundary (i.e.~not necessarily the ``Hartree-Fock point''). Its stability is the main result of our work and is explored in Secs.~\ref{sec:BDsetting}, \ref{sec:largersettings}. Consider a generalized Pauli constraint $D(\vec{\lambda})\geq 0$ (recall Eq.~(\ref{gPC}))
and for a given $N$-fermion state $|\Psi_N\rangle$ introduce the corresponding operator $\hat{D}_{\Psi_N}$ (according to Sec.~\ref{sec:notation}, where we now add an index ``$\Psi_N$'' for clarity),
\begin{equation}\label{Dop}
\hat{D}_{\Psi_N} = \kappa^{(0)} \mathds{1}+\kappa^{(1)} a_1^{\dagger} a_1+\ldots+\kappa^{(d)} a_{d}^{\dagger}a_{d}\,.
\end{equation}
Note that since $\kappa^{(i)} \in \mathbb{Z}$ we have $\mbox{spec}(\hat{D}_{\Psi_N}) \subset \mathbb{Z}$. $\hat{D}_{\Psi_N}$ is not positive semi-definite. Let $P_{\Psi_N}^{(\pm)}$ denote the projection operator onto the positive/negative eigenspace of $\hat{D}_{\Psi_N}$. Assume for a moment that $\hat{D}_{\Psi_N}$ is positive semi-definite. In that case if the NON $\vec{\lambda}_0$ of $|\Psi_N\rangle$ are pinned by $D(\vec{\lambda})\geq 0$, we find (recall (\ref{Functexpect}))
\begin{eqnarray}
0 &=& D(\vec{\lambda}_0) \nonumber \\
&=& \langle \Psi_N|\hat{D}_{\Psi_N}|\Psi_N\rangle\nonumber \\
&=&\langle \Psi_N|P_{\Psi_N}^{(+)} \hat{D}_{\Psi_N} P_{\Psi_N}^{(+)}|\Psi_N\rangle \nonumber \\
&\geq& \|P_{\Psi_N}^{(+)} \Psi_N \|_{L^2}^2\,,
\end{eqnarray}
where we used in the last step that $\mbox{spec}(\hat{D}_{\Psi_N}) \subset \mathbb{Z}$ and $\|\cdot\|_{L^2}$ is the $L^2$-norm.
This implies $\|P_{\Psi_N}^{(+)} \Psi_N \|_{L^2}=0$ and that $|\Psi_N\rangle$ only has weight in the $0$-eigenspace.
However, since $\hat{D}_{\Psi_N}$ is not positive semi-definite it seems in principle possible to have pinning, $D(\vec{\lambda}_0)=0$, with non-zero weights of $|\Psi_N\rangle$ in the positive and negative eigenspaces of $\hat{D}_{\Psi_N}$. In that case those two weights should cancel each other in the expression for $D(\vec{\lambda})$,
\begin{eqnarray}
D(\vec{\lambda}) &=& \langle\Psi_N|\hat{D}_{\Psi_N}|\Psi_N\rangle \\
&=& \underbrace{\langle\Psi_N| P_{\Psi_N}^{(-)} \hat{D} P_{\Psi_N}^{(-)}|\Psi_N\rangle}_{\leq 0}+
\underbrace{\langle\Psi_N| P_{\Psi_N}^{(+)} \hat{D}_{\Psi_N} P_{\Psi_N}^{(+)}|\Psi_N\rangle}_{\geq 0}\,. \nonumber
\end{eqnarray}

However, one finds the surprising result which was already implicitly stated in \cite{Kly1}.
\begin{thm}\label{lem:pin1}
Given $|\Psi_N\rangle \in \wedge^N[\mathcal{H}_1^{(d)}]$ with non-degenerate NON $\vec{\lambda}$ pinned by a gPC $D(\cdot)\geq 0$.
Then, $|\Psi_N\rangle$ has weight only in the $0$-eigenspace of $\hat{D}_{\Psi_N}$,
\begin{equation}\label{DPsi0}
\hat{D}_{\Psi_N}|\Psi_N\rangle  = 0\,.
\end{equation}
Equivalently, the self-consistent expansion (\ref{Psic}), $|\Psi_N\rangle = \sum_{\textbf{i}}\, c_{\textbf{i}}\,|\textbf{i}\,\rangle$, obeys the selection rule for Slater determinants
\begin{equation}\label{selrul}
\hat{D}_{\Psi_N} |\textbf{i}\,\rangle \neq 0 \qquad \Rightarrow \qquad c_{\textbf{i}} = 0\,.
\end{equation}
\end{thm}
\noindent Note that $\hat{D}_{\Psi_N} |\bd{i}\,\rangle$ can easily be calculated since the Slater determinants $|\bd{i}\,\rangle$ are the eigenstates of $\hat{D}_{\Psi_N}$ with eigenvalues $d_{\bd{i}} \in \mathbb{Z}$. Hence, $d_{\bd{i}}$ follows immediately from
\begin{equation}
d_{\bd{i}} = \langle \bd{i}\,| \hat{D}_{\Psi_N}|\bd{i}\,\rangle = \kappa^{(0)} + \sum_{k=1}^d\,\kappa^{(k)} \,\delta[k \in \bd{i}\,]\,,
\end{equation}
where $\delta[k \in \bd{i}\,]$ yields $1$  if $k \in \bd{i}$ and $0$ otherwise. This selection rule impressively shows the strength of the self-consistent notation (\ref{Psic}).

The proof of Theorem \ref{lem:pin1} is by far not elementary but rather involved. It can be performed adopting concepts like \emph{moment maps} used in symplectic geometry (see Lemma 2.1.~in Ref.~(\cite{Guillemin})).

To emphasize the importance of Theorem \ref{lem:pin1} and Selection Rule (\ref{selrul}) we study an example.
\begin{ex}\label{ex:BDpin}
Consider a state $|\Psi_3\rangle \in \wedge^3[\mathcal{H}_1^{(6)}]$ with non-degenerate NON $\vec{\lambda}$. The gPC are given by (\ref{d=6a}) and (\ref{d=6b}). The gPC (\ref{d=6a}) can be interpreted as inequalities, which are always saturated \cite{Kly3}. Consequently, Selection Rule (\ref{selrul}) implies structural simplifications for any $|\Psi_3\rangle \in \wedge^3[\mathcal{H}_1^{(6)}]$ expanded self-consistently,
\begin{equation}
|\Psi_3\rangle = \sum_{1\leq i_1 <i_2 < i_3\leq 6} c_{i_1,i_2,i_3}\,|i_1,i_2,i_3\rangle\,.
\end{equation}
Then only those Slater determinants $|i_1,i_2,i_3\rangle$ with exactly one index $i_k$ in each of the three sets $\{1,6\}$, $\{2,5\}$ and $\{3,4\}$ can show up in the expansion. These are the $2^3=8$ Slater determinants $|1,2,3\rangle$, $|1,2,4\rangle$, $|1,3,5\rangle$, $|1,4,5\rangle$, $|2,3,6\rangle$, $|2,4,6\rangle$, $|3,5,6\rangle$ and $|4,5,6\rangle$. Note that this structural simplification is not in contradiction with the dimension $\binom{6}{3}=20$ of the $3$-fermion Hilbert space $\wedge^3[\mathcal{H}_1^{(6)}]$, since the six $1$-particle states $\{|j\rangle\}_{j=1}^6$ depend on $|\Psi_3\rangle$.

If in addition  the NON $\vec{\lambda}$ are also pinned by the gPC (\ref{d=6b}), Selection Rule (\ref{selrul}) leads to the most general form
\begin{equation}\label{DBPinningstructure}
|\Psi_3\rangle = \alpha |1,2,3\rangle +\beta |1,4,5\rangle + \gamma |2,4,6\rangle\,.
\end{equation}
The three coefficients are such that
\begin{equation}
\lambda_1 = |\alpha|^2+|\beta|^2\,,\,\,\,\lambda_2 = |\alpha|^2+|\gamma|^2\,,\,\,\,\lambda_3 = |\alpha|^2\,.
\end{equation}
\end{ex}
This example demonstrates the strength of the self-consistent expansion (\ref{Psic}) and the significant reduction from twenty to three Slater determinants spanning $|\Psi_3\rangle \in \wedge^3[\mathcal{H}_1^{(6)}]$, in case of pinning.

For the later purpose we define
\begin{defn}\label{def:projD=0}
For the setting $\wedge^N[\mathcal{H}_1^{(d)}]$, a corresponding gPC $D(\vec{\lambda})\geq 0$ and a given state $|\Psi_N\rangle \in \wedge^N[\mathcal{H}_1^{(d)}]$ with non-degenerate NON and NO $\{|i\rangle\}_{i=1}^d$ consider the corresponding operator $\hat{D}_{\Psi_N}$
(recall Eq.~(\ref{Fop})). We denote the orthogonal projection operator on the $0$-eigenspace of $\hat{D}_{\Psi_N}$ by $P_{D}$. $P_{D}$ projects onto
the linear space spanned by all those Slater determinants $|\textbf{i}\,\rangle$, which fulfill Selection Rule (\ref{selrul}).
\end{defn}

Note also that for a gPC (\ref{gPC}) with $\kappa^{(i)}=\kappa^{(j)}$ we find $P_D = P_{\pi_{i,j} D}$, where $\pi_{i,j}$ just swaps the $i$-th ($\lambda_i$) and $j$-th argument ($\lambda_j$) of $D(\cdot)$. This means that the family of Slater determinants $|k_1,\ldots,k_N\rangle$ spanning the $P_D$-subspace is invariant under swapping the indices $i$ and $j$ in all Slater determinants.

Let us summarize all insights up to now.
\begin{rem}\label{rem:Pinningisrelev}
Pinning corresponds to specific and simplified structures of the corresponding $N$-fermion quantum state $|\Psi_N\rangle$. In that sense the occurrence of pinning is physically relevant. It is also remarkable that pinning as phenomenon in the $1$-particle picture allows one to reconstruct the structure of $|\Psi_N\rangle$ as object in the important $N$-particle picture.
\end{rem}

Theorem \ref{lem:pin1} and Remark \ref{rem:Pinningisrelev} provide a first insight into whether pinning may exist in realistic physical systems:
\begin{rem}\label{rem:PinArtificial}
Consider a physical system of a few interacting fermions in a continuous space confined by some external potential without any exact symmetry. By expressing any state $|\Psi_N\rangle$ of that system (e.g.~the ground state) as a linear combination of Slater determinants built up from its NO according to (\ref{Psic}) it is unlikely that some (or even almost all) expansion coefficients do \emph{exactly} vanish. As a consequence of Theorem \ref{lem:pin1} and Selection Rule (\ref{selrul}) therein pinning may, in fact, not show up for such realistic physical systems. However, if the system has certain symmetries as e.g.~rotational or translational invariance manifesting itself in the existence of  good quantum numbers $Q$ as e.g.~the total angular momentum or the total wave number of interacting electrons only Slater determinants with the same quantum numbers $Q$ do show up in the expansion of $|\Psi_N\rangle$. In that case
at least, it seems more feasible, in principle, to have pinning.
\end{rem}
According to Remark \ref{rem:PinArtificial} pinning in its idealized exact form may not show up for realistic fermionic systems.
However, since systems of few fermions in a steep external potential as e.g.~a harmonic trap have the strong tendency to minimize their energy
a few specific $N$-fermion configurations are strongly preferred. For the expansion (\ref{Psic}) of the ground state $|\Psi_N\rangle$ this means that all except just a few weights $|c_{\bd{i}}|^2$ are quite small. By recalling that
\begin{equation}
D(\vec{\lambda}) = \langle \Psi_N| \hat{D}_{\Psi_N}|\Psi_N\rangle
\end{equation}
depends continuously on $|\Psi_N\rangle$ one cannot rule out that some generalized Pauli constraints are at least \emph{approximately} saturated \footnote{That $|\Psi_N\rangle$ is (approximately) spanned by just a few Slater determinants of course does not imply at all that its NON are (quasi)pinned by some gPC $D(\vec{\lambda})\geq 0$ but it is a necessary condition for it according to Theorem \ref{lem:pin1} and Selection Rule (\ref{selrul}) therein.}. For the ground state of a system of three harmonically coupled fermions in a harmonic trap this effect of \emph{quasipinning} rather than pinning was found analytically \cite{CS2013}.

According to those first analytic results and Remark \ref{rem:PinArtificial} the central question is whether Theorem \ref{lem:pin1} is stable: Do NON $\vec{\lambda}$ in the vicinity of some polytope facet $F_D$ imply that the corresponding $N$-fermion state $|\Psi_N\rangle$ has approximately the specific and simplified structure expressed by Selection Rule (\ref{selrul})?

Unfortunately, such stability of Selection Rule (\ref{selrul}) is not given for the whole vicinity of the polytope boundary as the following example shows.
\begin{ex}\label{ex:unstable}
For the Borland-Dennis setting $\wedge^3[\mathcal{H}_1^{(6)}]$ with the gPC (\ref{d=6a}), (\ref{d=6b}) consider a state of the form
\begin{equation}\label{DBcounterex}
|\Psi_3\rangle = \gamma \, |1,3,5\rangle +\sqrt{|\gamma|^2+|\delta|^2-\varepsilon} \,\,|1,2,4\rangle + \delta \,|2,3,6\rangle\,.
\end{equation}
with $|\gamma| > |\delta|$ and $\varepsilon>0$. State (\ref{DBcounterex}) is indeed self-consistent since its NO are
given by $\{|j\rangle\}_{j=1}^6$ and their occupancies $\lambda_j$ are non-degenerate and decreasingly ordered. The approximate saturation of the gPC (\ref{d=6b}) is $D^{(3,6)}(\vec{\lambda}) = \varepsilon$, which can be made arbitrarily small by choosing  $\varepsilon$ in Eq.~(\ref{DBcounterex}) arbitrarily small.
However, state (\ref{DBcounterex}) maximally contradicts a possible stability of Selection Rule (\ref{selrul}). None of the three suggested Slater determinants in Ex.~\ref{ex:BDpin} shows up. However, state (\ref{DBcounterex}) has the specific property that possible strong quasipinning (i.e.~$\varepsilon \ll 1$ ) is equivalent to an approximate saturation of the ordering constraints $\lambda_3-\lambda_4 \geq 0$ since $\lambda_3-\lambda_4 = \varepsilon$, i.e.~both NON become degenerate.
\end{ex}

The specific instability of Selection Rule (\ref{selrul}) shown by Ex.~\ref{ex:unstable} is  not surprising, at least from a heuristic viewpoint.
First, notice that the three Slater determinants $|1,3,5\rangle$, $|1,2,4\rangle$ and $|2,3,6\rangle$ in Eq.(\ref{DBcounterex}) are related to the three
Slater determinants $|1,4,5\rangle$, $|1,2,3\rangle$ and $|2,4,6\rangle$ in Eq.(\ref{DBPinningstructure}) corresponding to pinning of non-degenerate NON (Ex.~\ref{ex:BDpin}) by a swapping of the integers $3$ and $4$. Second, the self-consistent expansion (\ref{Psic}) to which Selection Rule (\ref{selrul}) refers to is based on ordered NO $\{|i\rangle\}$. However, if two NON $\lambda_i$ and $\lambda_{i+1}$ are identical the choice of NO $|i\rangle$ and $|i+1\rangle$ is not unique anymore. In that sense it is not surprising that for the case of quasidegenerate NON with $\lambda_i\approx \lambda_{i+1}$
not only Slater determinants $|\bd{j}\,\rangle$ lying in the $P_D$-subspace (recall Def.~\ref{def:projD=0}) can show up in the self-consistent expansion (\ref{Psic}) but also those obtained by swapping the integers $i$ and $i+1$.  That this is indeed a correct modification of Selection Rule (\ref{selrul}) for the case of degenerate and pinned NON is strongly suggested by our following results on quasipinning and its structural implications.

\section{Quasipinning and stability of Selection Rule (\ref{selrul})}\label{sec:quasipinning}
The discussion in Sec.~\ref{sec:physrev} has shown that the NON of an energy eigenstate $|\Psi_N\rangle$ of a realistic model hamiltonian may not be pinned by gPC since $|\Psi_N\rangle$ may not fulfill the quite restrictive Selection Rule (\ref{selrul}) for Slater determinants in Theorem \ref{lem:pin1}.
More realistic is the effect of quasipinning which was found for a model system \cite{CS2013}. However, the occurrence of this effect is only physically relevant when Selection Rule (\ref{selrul}) still applies at least approximately. To investigate the possible stability of Selection Rule (\ref{selrul}) is the goal of this section. We also quantify the stability in particular for the regime
characterized by an additional approximate saturation of an ordering constraint $\lambda_i-\lambda_{i+1}\geq 0$. 

We start by providing a lower bound on the stability for all $N$ and $d$. The following result states quantitatively that whenever given NON $\vec{\lambda}$ are close but not exactly on a polytope facet $F_D$ any corresponding state $|\Psi_N\rangle$ has at least some weight outside the $P_D$-subspace:
\begin{thm}\label{stat:ConverseStab}
Recall Def.~\ref{def:projD=0} and Eq.~(\ref{Dop}). For NON $\vec{\lambda}$ belonging to the setting $\wedge^N[\mathcal{H}_1^{(d)}]$ and a gPC $D(\cdot)\geq 0$ we have
\begin{equation}
1-\|P_D \Psi_N\|_{L^2}^2 \geq \frac{D(\vec{\lambda})}{\|\hat{D}_{\Psi_N}\|_{\small{\mbox{op}}}}\,,
\end{equation}
where $\|\hat{D}_{\Psi_N}\|_{\small{\mbox{op}}}$ is the operator norm.
\end{thm}
The proof is elementary and is shown in Appendix \ref{app:QPlowerbound}.
The following example gives an illustration of Theorem \ref{stat:ConverseStab}
\begin{ex}\label{ex:Converse}
For the gPC (\ref{d=6b}) and (\ref{gPC37}) of the setting $\wedge^3[\mathcal{H}_1^{(6)}]$ and $\wedge^3[\mathcal{H}_1^{(7)}]$, respectively,
the maximal eigenvalue $\|\hat{D}_{\Psi_3}\|_{\small{\mbox{op}}}$ of the corresponding $\hat{D}_{\Psi_3}$-operator is always given by $2$. Then Theorem \ref{stat:ConverseStab} provides for each gPC of those two settings the following lower bound on stability
\begin{equation}
1-\|P_D \Psi_3\|_{L^2}^2   \geq \frac{1}{2}\,D(\vec{\lambda}) \,.
\end{equation}
\end{ex}

\subsection{Borland-Dennis setting}\label{sec:BDsetting}
The Borland-Dennis setting is the ``smallest'' setting with nontrivial gPC (cf.~Eqs.~(\ref{d=6a}) and (\ref{d=6b})).
Since on the other hand the number of gPC increases drastically for increasing $N$ and $d$ it is due to the
manageable structure of its polytope $\mathcal{P}_{3,6}$ the most relevant setting for physical applications.
Potentially relevant might be also the setting $\wedge^3[\mathcal{H}_1^{(7)}]$ whose polytope is described by four gPC and is studied in Sec.~\ref{sec:setting37}. Since already the next larger settings $\wedge^3[\mathcal{H}_1^{(8)}]$, $\wedge^4[\mathcal{H}_1^{(8)}]$, $\wedge^3[\mathcal{H}_1^{(9)}]$ and $\wedge^4[\mathcal{H}_1^{(9)}]$ are described by $31$, $14$, $51$ and $59$ gPC, respectively, their potential relevance for physical application is not obvious. E.g.~that a given vector $\vec{\lambda}$ of NON is pinned to the polytope boundary of such a setting is useless without the specification of the concrete facet which $\vec{\lambda}$ lies on. Even more important, the central question of the mechanism behind quasipinning needs to be addressed for every gPC separately and therefore seems to make less sense for settings with more than just a very few gPC.

As a consequence, we explore the possible stability of Selection Rule (\ref{selrul}) for the  Borland-Dennis setting first.
This will be done analytically and numerically.

For any given state $|\Psi_3\rangle \in \wedge^3[\mathcal{H}_1^{(6)}]$ conditions (\ref{d=6a}) imply according to Ex.~\ref{ex:BDpin}
structural simplifications:
\begin{eqnarray}\label{8SD}
|\Psi_3\rangle &=& \alpha |1,2,3\rangle+ \beta |1,2,4\rangle+ \gamma |1,3,5\rangle\nonumber \\
&& + \delta |2,3,6\rangle +\nu |1,4,5\rangle+\mu |2,4,6\rangle \nonumber \\
&&+ \xi |3,5,6\rangle+\zeta |4,5,6\rangle \,.
\end{eqnarray}
The eight coefficients $\alpha,\ldots,\zeta$ obey further restrictions, guaranteeing that the NON are decreasingly ordered and that $\rho_1$ is diagonal
w.r.t~the NO $|k\rangle,  k=1,2,...6$. Consequently, we have
\begin{eqnarray}
\lambda_4 &=& |\beta|^2+|\mu|^2+|\nu|^2+|\zeta|^2   \nonumber \\
\lambda_5 &=& |\gamma|^2+|\nu|^2+|\xi|^2+|\zeta|^2  \nonumber \\
\lambda_6 &=& |\delta|^2+|\mu|^2+|\xi|^2+|\zeta|^2\,
\end{eqnarray}
and the largest three NON follow from Eq.~(\ref{d=6a}).
In the following we will use the geometric picture introduced in Sec.~\ref{sec:notation} (recall in particular Eqs.~(\ref{Paulicube}), (\ref{NONgeom})).
Due to the three conditions (\ref{d=6a}) the NON are not independent and we choose $\lambda_4,\lambda_5$ and $\lambda_6$ as the free
variables and the polytope $\mathcal{P}_{3,6}$ reduces to a polytope $\hat{\mathcal{P}}_{3,6}$ of possible vectors $\vec{v}\equiv (\lambda_4,\lambda_5,\lambda_6) \subset \mathbb{R}^3$.
\begin{figure}[h]
\centering
\includegraphics[width=0.4\textwidth]{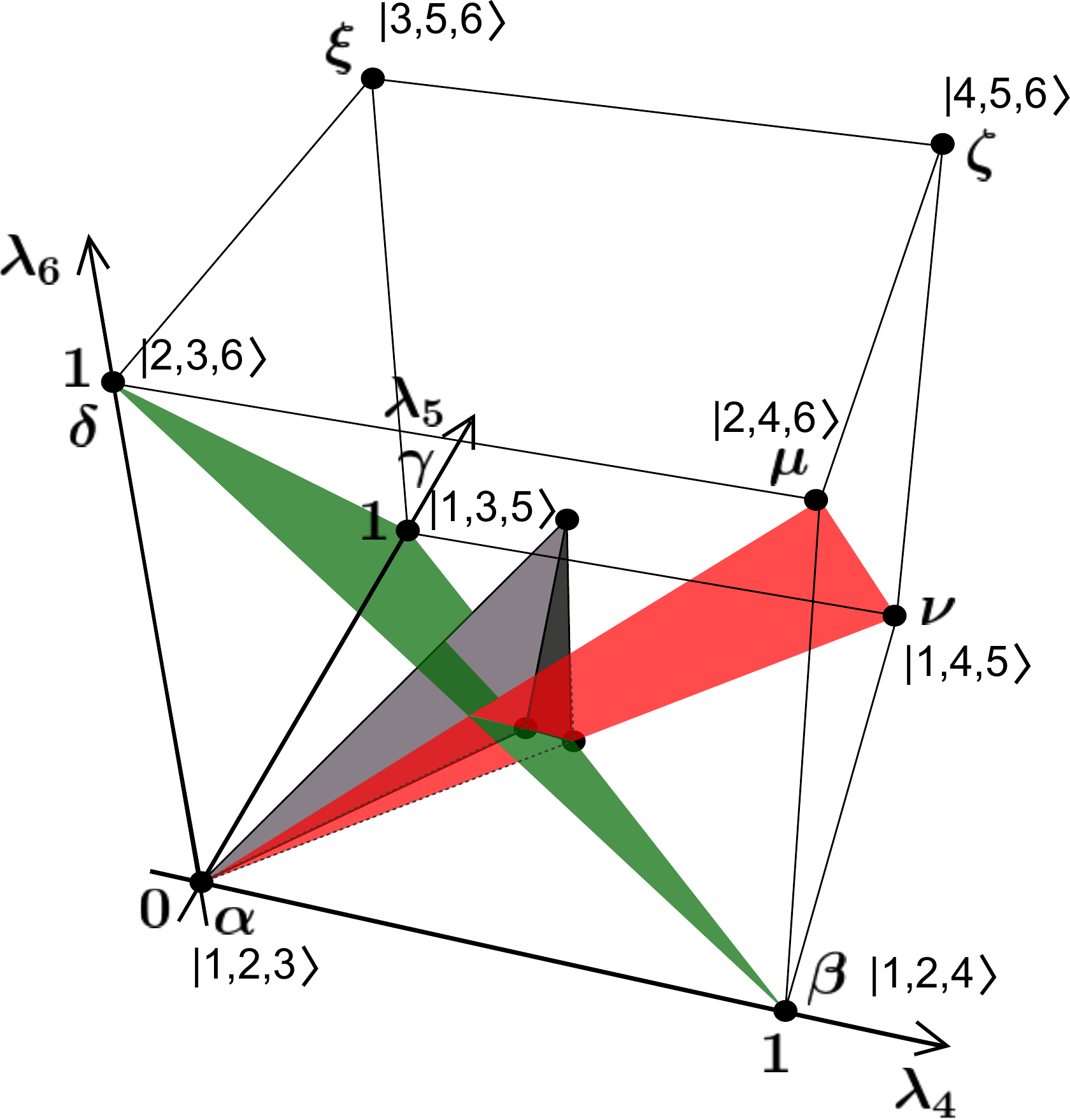}
\captionC{Reduced polytope $\hat{\mathcal{P}}_{3,6}$ (gray) of possible (independent) NON $\vec{v}\equiv(\lambda_4,\lambda_5,\lambda_6)$, where one of its four vertices lies at the center, $(\frac{1}{2},\frac{1}{2},\frac{1}{2})$, of the cube $[0,1]^3$. For a detailed explanation see text.
}
\label{fig:polytope}
\end{figure}
This reduced polytope $\hat{\mathcal{P}}_{3,6}$ is shown in Fig.~\ref{fig:polytope}, and is spanned by the four vertices
\small \begin{eqnarray}
&&\vec{v}^{(a)}= \left( 0,0,0\right)\,,\,\, \vec{v}^{(b)}= \left( \frac{1}{2}\ \frac{1}{2},0\right) \nonumber \\
&& \vec{v}^{(c)}= \left( \frac{1}{2} , \frac{1}{4} ,\frac{1}{4} \right) \,,\,\,  \vec{v}^{(d)}= \left( \frac{1}{2} , \frac{1}{2} ,\frac{1}{2}\right)\,.
\end{eqnarray}
\normalsize
Due to the ordering constraints, $\frac{1}{2} \geq \lambda_4\geq \lambda_5 \geq \lambda_6\geq 0$, $\hat{\mathcal{P}}_{3,6}$ lies in the lower, left, front
octant. Fig.~\ref{fig:polytope} also shows the strength of the geometric picture. Each (reduced) vector $\vec{v}=(\lambda_4,\lambda_5,\lambda_6)$ is the ``center of mass'' of the $8$ ``masses'' $|\alpha|^2,\ldots,|\zeta|^2$ sitting at the
eight vertices of the (reduced) Pauli cube $[0,1]^3$. These vertices $\vec{v}_{\bd{i}}$ can be identified with the eight Slater determinants $|\bd{i}\,\rangle$ used in the expansion (\ref{8SD}), since their occupancies $(\langle \bd{i}\,|\hat{n}_j|\bd{i}\,\rangle)_{j=1}^6$ w.r.t.~$\{|j\rangle\}_{j=1}^6$ coincide with $\vec{v}_{\bd{i}}$. Therefore, we assigned the Slater determinants $|\bd{i}\,\rangle$ as well as the corresponding Greek letters $\alpha,\ldots,\zeta$
to them in Fig.~\ref{fig:polytope}. The eight weights $|\alpha|^2,\ldots,|\zeta|^2$ need to fulfill the self-consistency condition $\vec{v} \in \hat{\mathcal{P}}_{3,6}$. Pinning by gPC (\ref{d=6b}) is given whenever $\vec{v} \in \hat{\mathcal{P}}_{3,6}$ lies on the (red) plane which is spanned by the points $\alpha,\mu$ and $\nu$. The distance of arbitrary NON vectors $\vec{v}$ to the (red) plane is a measure for the strength of the possible quasipinning by gPC (\ref{d=6b}). In general, since the polytopes $P_{N,d}$ are finite-dimensional all norms are equivalent. In particular it can be seen (see e.g.~\cite{CSthesis}) that the distances to some facet $F_{D}$ corresponding to a gPC $D(\vec{\lambda}) \geq 0$ are given by
\begin{equation}
\mbox{dist}_2(\vec{\lambda},F_{D}) = \frac{D(\vec{\lambda})}{\|\vec{\kappa}\|_2} \,,\,\mbox{dist}_1(\vec{\lambda},F_{D}) = \frac{D(\vec{\lambda})}{\|\vec{\kappa}\|_{\infty}}\,,
\end{equation}
where
$\mbox{dist}_2$ and $\|\vec{\kappa}\|_2$ are the $l^2$/Euclidean-distance, $\mbox{dist}_1$ the $l^1$-distance, $\|\vec{\kappa}\|_{\infty}$
the max norm and $\vec{\kappa}$ from Eq.~(\ref{gPC}) is a normal vector of $F_{D}$. For the gPC (\ref{d=6b}) of the Borland-Dennis setting we find
\begin{eqnarray}
\mbox{dist}_2(\vec{\lambda},F_{D^{(3,6)}}) &=& \frac{D^{(3,6)}(\vec{\lambda})}{\sqrt{3}} \nonumber \\
\mbox{dist}_1(\vec{\lambda},F_{D^{(3,6)}}) &=& D^{(3,6)}(\vec{\lambda})\,.
\end{eqnarray}
Therefore, we just choose $D(\vec{\lambda})$ as the natural measure for quasipinning by gPC $D(\vec{\lambda})\geq 0$.

Now, we explore the stability of Selection Rule (\ref{selrul}). For non-degenerate NON $\vec{\lambda} \in F_{D^{(3,6)}}$, according to Ex.~\ref{ex:BDpin}, $|\Psi_3\rangle$ is spanned just by the three points $\alpha,\mu$ and $\nu$, the vertices of the (red) plane in Fig.~\ref{fig:polytope}. For quasipinning the stability can be violated as was shown by Ex.~\ref{ex:unstable}. There, $|\Psi_3\rangle$ was spanned by the points $\beta, \gamma$ and $\delta$, which span the (green) plane, defined by
\begin{equation}\label{BDplaneDalt}
Q(\lambda_4,\lambda_5,\lambda_6) = D^{(3,6)}(1-\lambda_4,\lambda_5,\lambda_6) =0\,
\end{equation}
where we skipped the first three variables $\lambda_1,\lambda_2,\lambda_3$ and treat from now on $D^{(3,6)}$ as function of just $\vec{v}=(\lambda_4,\lambda_5,\lambda_6)$.
The (green) plane is related to the (red) plane
\begin{equation}\label{BDplaneD}
D^{(3,6)}(\lambda_4,\lambda_5,\lambda_6) =0
\end{equation}
by a swapping of $\lambda_3$ and $\lambda_4$, i.e.~by replacing $\lambda_4$ by $1-\lambda_4$ (recall gPC (\ref{d=6a})).
Geometrically, we easily see that a strong quasipinning for the state in Ex.~\ref{ex:unstable} with a distance $\varepsilon \ll 1$ to $F_{D^{(3,6)}}$ is equivalent to an approximate saturation of the ordering constraint $\Delta \lambda = \lambda_3-\lambda_4 \geq 0$ (i.e.~$\lambda_4\approx \frac{1}{2}$).
For such a regime, $\lambda_i\approx \lambda_{i+1}$, we suggested below Ex.~\ref{ex:unstable} a modified selection rule.
According to it not only Slater determinants lying in the $P_D$-subspace but also those in the $P_{\pi_{i,i+1}D}$-subspace can show up in the expansion (\ref{Psic}), where $\pi_{i,i+1}$ is the swap-operator for coordinate $\lambda_i$ and $\lambda_{i+1}$.  On the other hand, this also suggests that even if some ordering constraint $\lambda_i-\lambda_{i+1}\geq 0$ is (approximately) saturated all the Slater determinants neither lying in $P_D$- nor in the $P_{\pi_{i,i+1}D}$-subspace should not carry any significant weights.


The following analytic result provides first insights into the stability of Selection Rule (\ref{selrul}).
\begin{thm}\label{lem:xizeta}
For a given $|\Psi_3\rangle\in \wedge^3[\mathcal{H}_1^{(6)}]$ expanded according to Eq.~(\ref{8SD}) we have
\begin{equation}\label{xizetabound}
|\xi|^2+|\zeta|^2 \leq D^{(3,6)}(\vec{\lambda})\,.
\end{equation}
In particular, this means that possible quasipinning by (\ref{d=6b}) implies an approximate
structural simplification for $|\Psi_3\rangle$.
\end{thm}
The proof of this important theorem is presented in Appendix \ref{app:xizeta}.

To gain additional information and a potential improvement of the upper bound in Eq.~(\ref{xizetabound})
we explore the weight $|\xi|^2+|\zeta|^2$ numerically.
\begin{figure}[h]
\centering
\includegraphics[width=0.23\textwidth]{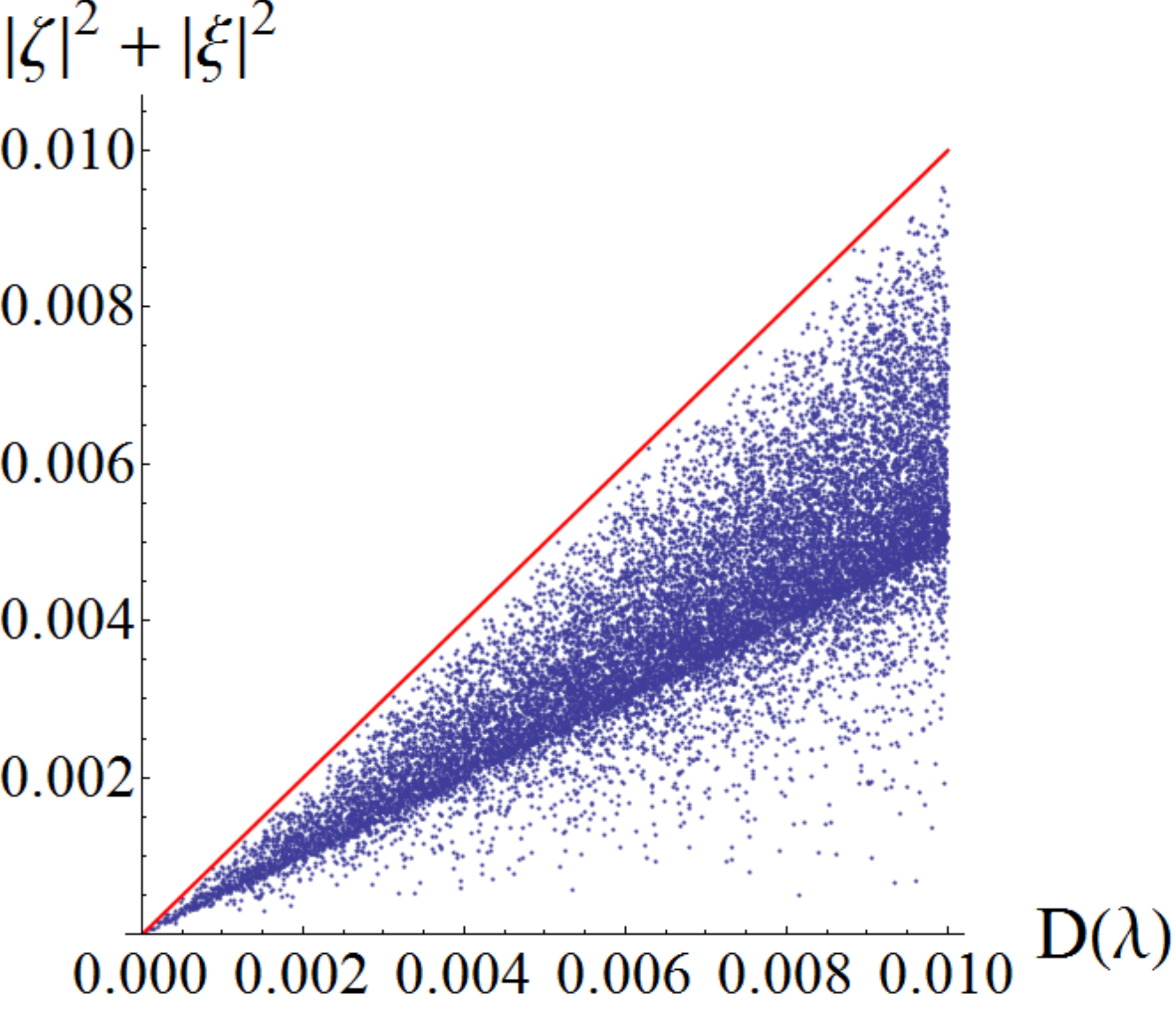}
\includegraphics[width=0.23\textwidth]{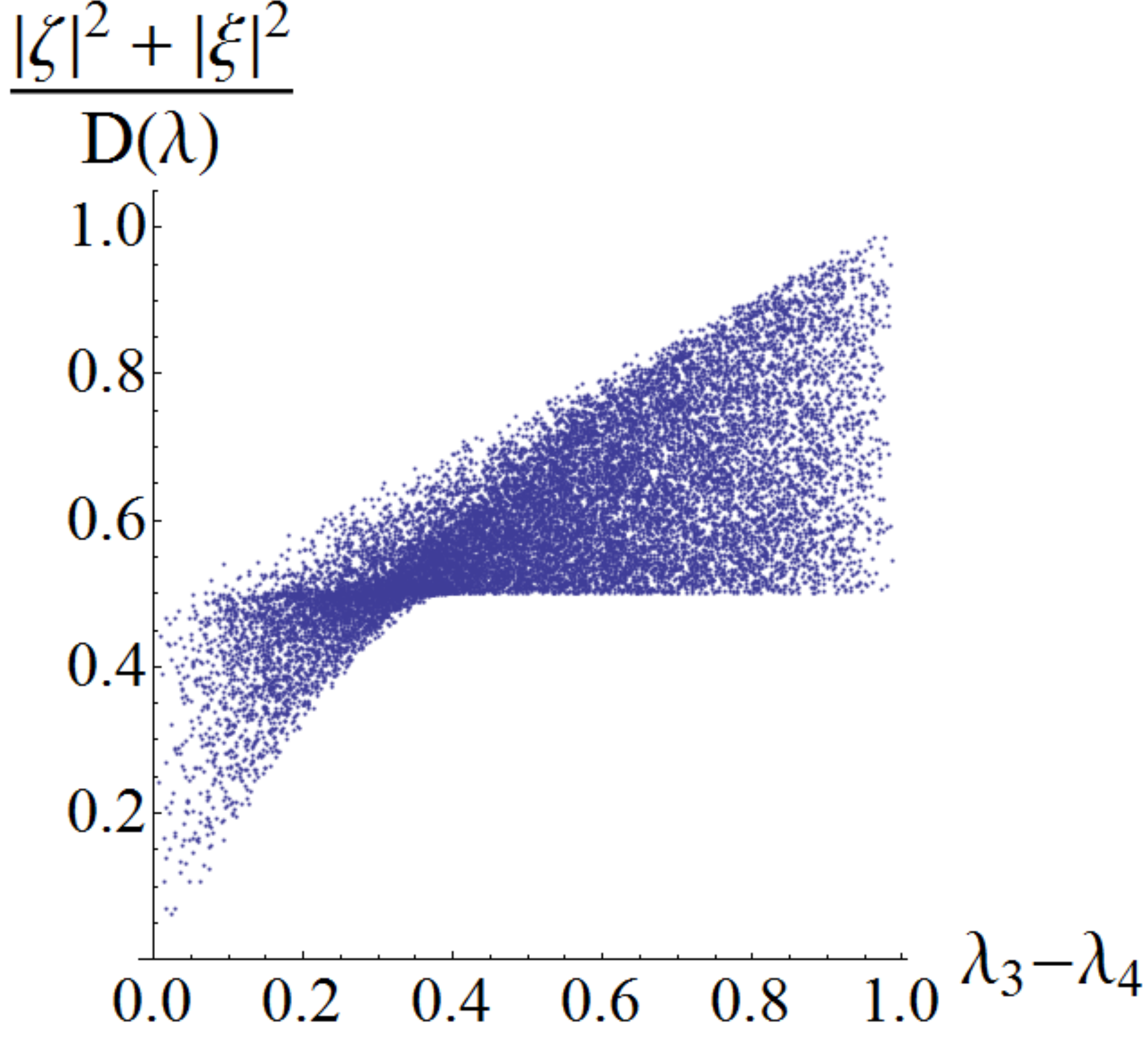}
\captionC{Sampling of states $|\Psi_3\rangle \in \wedge^3[\mathcal{H}_1^{(6)}]$ confirms Theorem \ref{lem:xizeta}. Left: $|\xi|^2+|\zeta|^2$ is bounded from above by $D^{(3,6)}(\vec{\lambda})$, the red straight line. Right: Even stronger bounds are revealed by a resolution w.r.t.~$\lambda_3-\lambda_4$.}
\label{fig:36xizeta}
\end{figure}
We sample random states $|\Psi_3\rangle \in \wedge^3[\mathcal{H}_1^{(6)}]$.
For all states exhibiting quasipinning, which we define here as $D^{(3,6)}(\vec{\lambda})\leq 0.01$ \footnote{The maximal possible $l^1$-distance $D^{(3,6)}(\vec{\lambda})$ of NON $\vec{\lambda}$ to the facet $F_{D^{(3,6)}}$ is $\frac{1}{2}$ and found for the NON $(\frac{1}{2},\frac{1}{2},\frac{1}{2},\frac{1}{2},\frac{1}{2},\frac{1}{2})$.} we calculate the corresponding  NO, NON and the expansion coefficients $\alpha,\ldots,\zeta$ for the self-consistent expansion (\ref{8SD}). The result of the sampling process is shown in Fig.~\ref{fig:36xizeta}.
On the left we can see that the weight $|\xi|^2+|\zeta|^2$ is indeed linearly bounded from above by $D^{(3,6)}(\vec{\lambda})$. The numerical results on the left do not change qualitatively by extending the range of the horizontal axis from the regime of strong quasipinning $D^{(3,6)}(\vec{\lambda})\leq 0.01$ to arbitrary values $D^{(3,6)}(\vec{\lambda})$. On the right we explore the tightness of the bound in Theorem \ref{lem:xizeta} as a function of $\lambda_3-\lambda_4$.
The reason for considering only the difference between $\lambda_3$ and $\lambda_4$ will become clearer at the end of this section (Sec.~\ref{sec:BDsetting}).
We can infer that bound $(\ref{xizetabound})$ is tight for a maximal distance between $\lambda_3$ and $\lambda_4$, which corresponds to the ``Hartree-Fock point'' $\vec{\lambda}=(1,1,1,0,0,0)$. For the regime $\lambda_3-\lambda_4\approx 0$ the estimate (\ref{xizetabound}) could be strengthened by reducing the upper bound $D^{(3,6)}(\vec{\lambda})$ by a factor $\frac{1}{2}$.

\begin{figure}[h]
\centering
\includegraphics[width=0.23\textwidth]{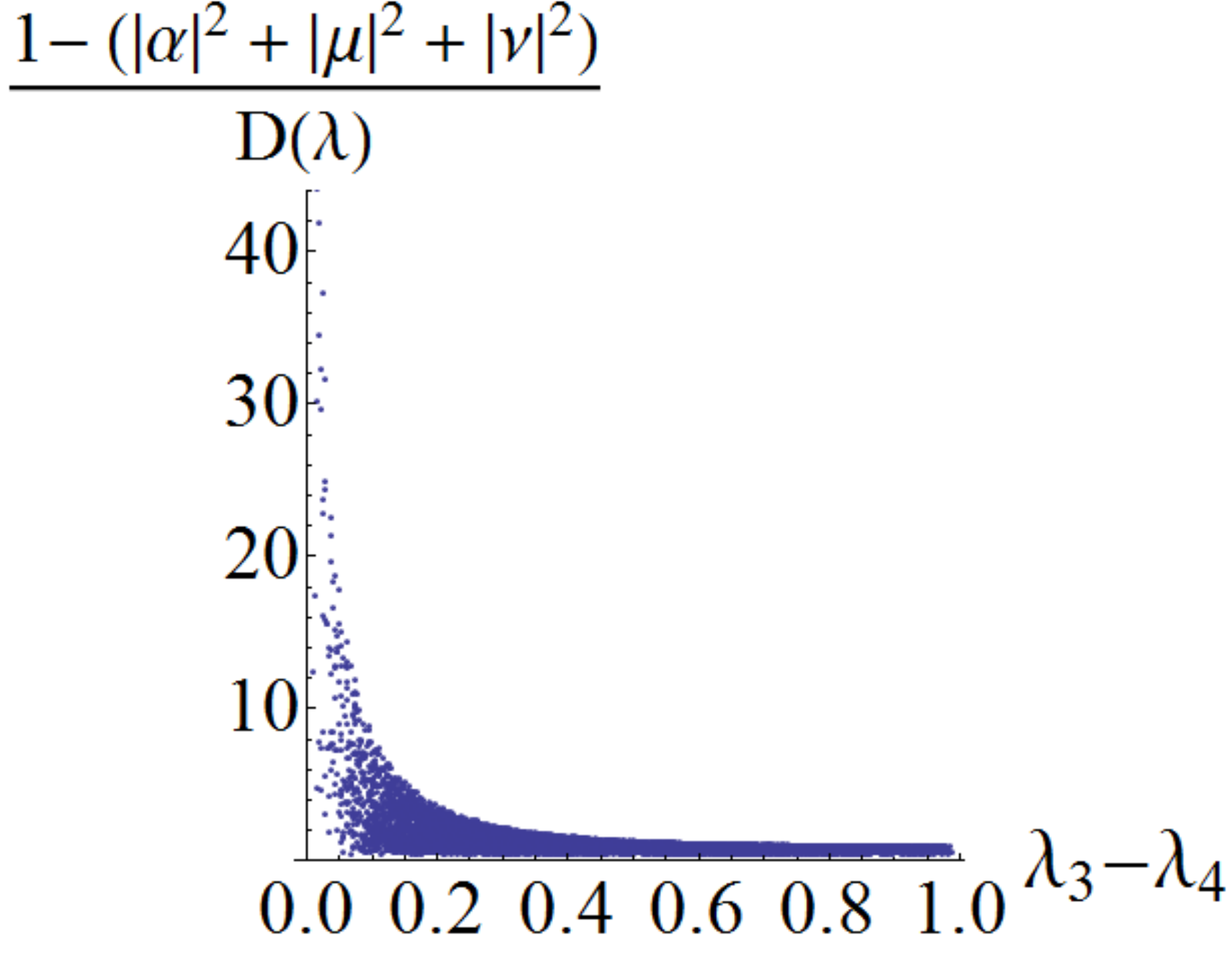}
\includegraphics[width=0.23\textwidth]{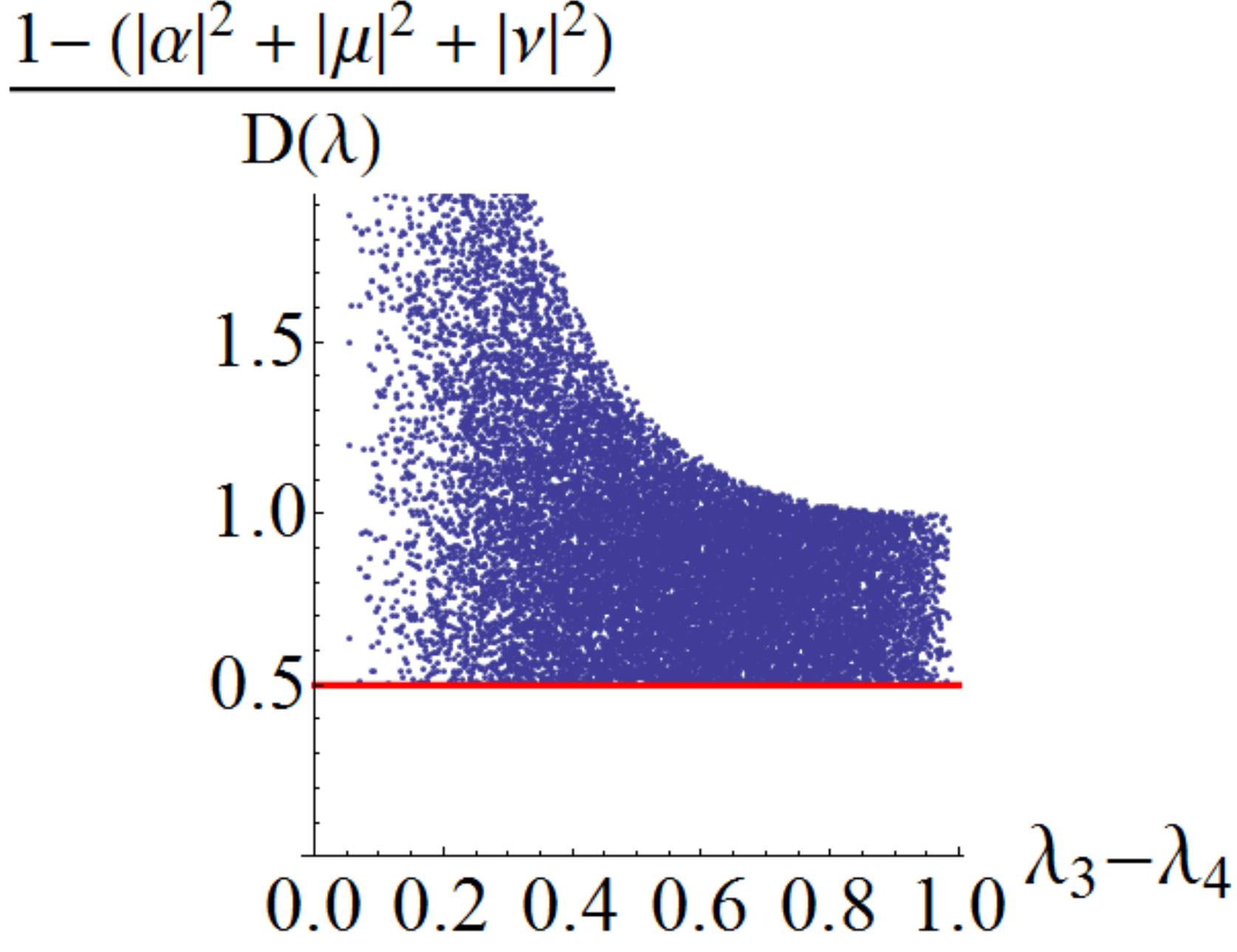}
\captionC{Sampling of states $|\Psi_3\rangle \in \wedge^3[\mathcal{H}_1^{(6)}]$ shows that Selection Rule (\ref{selrul}) is stable for the Borland-Dennis setting as long as the ordering constraint $\lambda_3-\lambda_4\geq 0$ is not approximately saturated (left). On the right: For fixed quasipinning $D^{(3,6)}(\vec{\lambda})$ there is a lower bound (red straight line) on the $L^2$-weight $1-(|\alpha|^2+|\mu|^2+|\nu|^2)$.}
\label{fig:36stable}
\end{figure}

Although Theorem \ref{lem:xizeta} is already quite a strong result on potential physical relevance of quasipinning it is still even not clear whether Selection Rule (\ref{selrul}) is stable whenever none of the ordering constraints is approximately saturated. The numerical investigation by sampling random states provides the answer. This is shown in Fig.~\ref{fig:36stable}.
On the left is shown the square of the $L^2$-weight (recall Def.~\ref{def:projD=0}) $1-\|P_{D^{(3,6)}}\Psi_3\|_{L^2}^2=1-\big(|\alpha|^2+|\mu|^2+|\nu|^2\big)$ divided by $D^{(3,6)}(\vec{\lambda})$ for randomly sampled states $|\Psi_3\rangle \in \wedge^3[\mathcal{H}_1^{(6)}]$. Selection Rule (\ref{selrul}) is stable whenever quasipinning implies that $\|P_D \Psi_3\|_{L^2}^2 \approx 1$. We can see that this is the case except for the regime $\lambda_3 \approx \lambda_4$.

On the right side of Fig.~\ref{fig:36stable} we can see that Selection Rule (\ref{selrul}) also applies in the converse direction. If $D^{(3,6)}(\vec{\lambda})$ is still finite not all weight can lie in the $P_{D^{(3,6)}}$-subspace. There, the \emph{lower} bound $\frac{1}{2}$ (red straight line) is in agreement with Ex.~\ref{ex:Converse}.

The numerical results shown in Fig.~\ref{fig:36stable} can also be understood analytically:
\begin{thm}\label{lem:unstablebound}
For given $|\Psi_3\rangle\in \wedge^3[\mathcal{H}_1^{(6)}]$ expanded according to Eq.~(\ref{8SD}) we find
\begin{equation}\label{unstablebound}
\|P_{\pi_{3,4} D} \Psi_3\|_{L^2}^2=|\beta|^2+|\gamma|^2+|\delta|^2  \leq \frac{D^{(3,6)}(\vec{\lambda})}{\lambda_3-\lambda_4}+3 D^{(3,6)}(\vec{\lambda})\,.
\end{equation}
\end{thm}
Theorem \ref{lem:unstablebound} explains that a violation of Selection Rule (\ref{selrul}) is only possible if the approximate saturation of $\lambda_3-\lambda_4\geq 0$ is at least as strong as the quasipinning by (\ref{d=6b}). This is in agreement with Ex.~\ref{ex:unstable}. There the violation Eq.(\ref{selrul}) was related to an approximate saturation of $\lambda_3-\lambda_4\geq 0$. The proof of Theorem \ref{lem:unstablebound} is presented in Appendix \ref{app:unstablebound}.
\begin{figure}[h]
\centering
\includegraphics[width=0.22\textwidth]{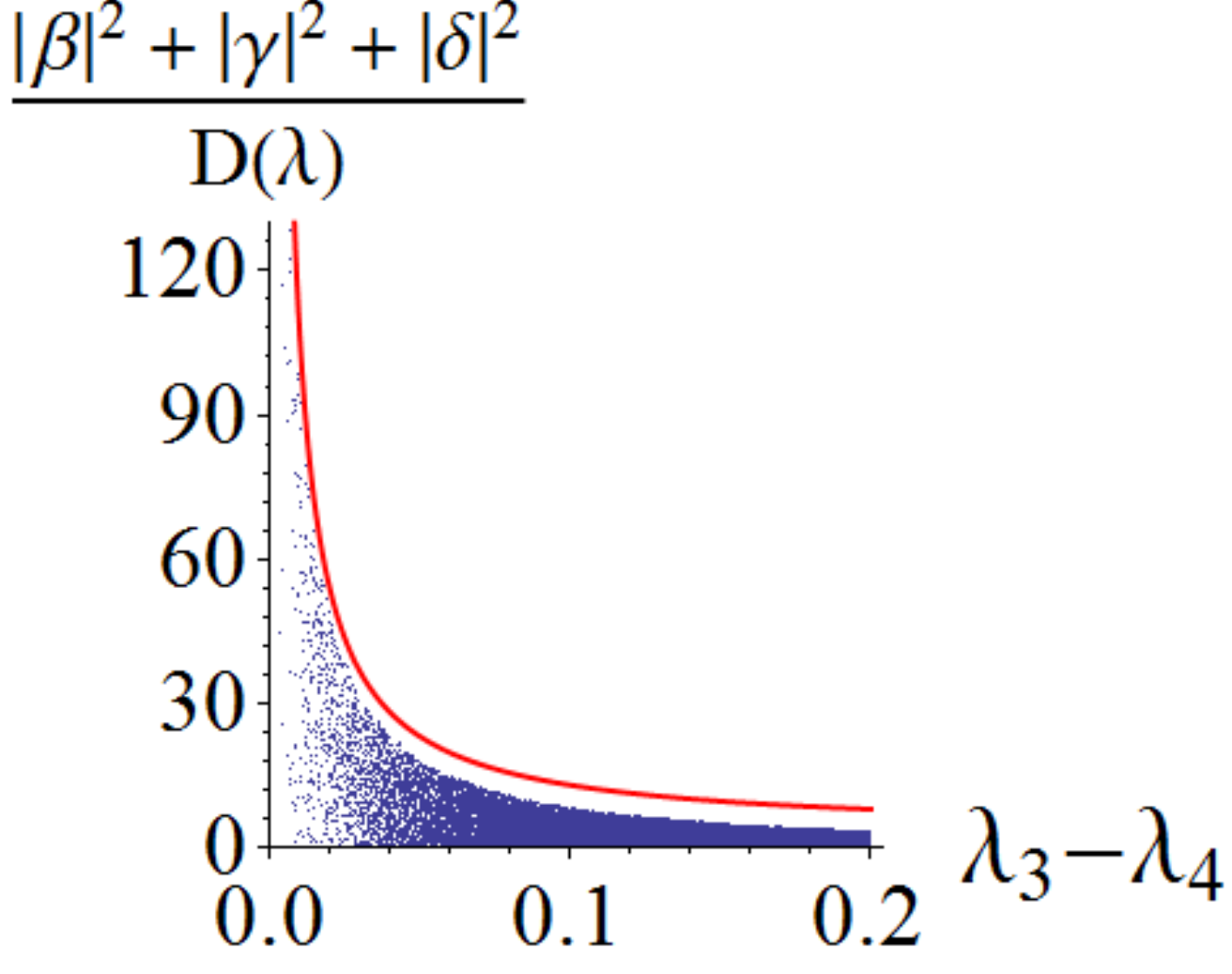}
\includegraphics[width=0.24\textwidth]{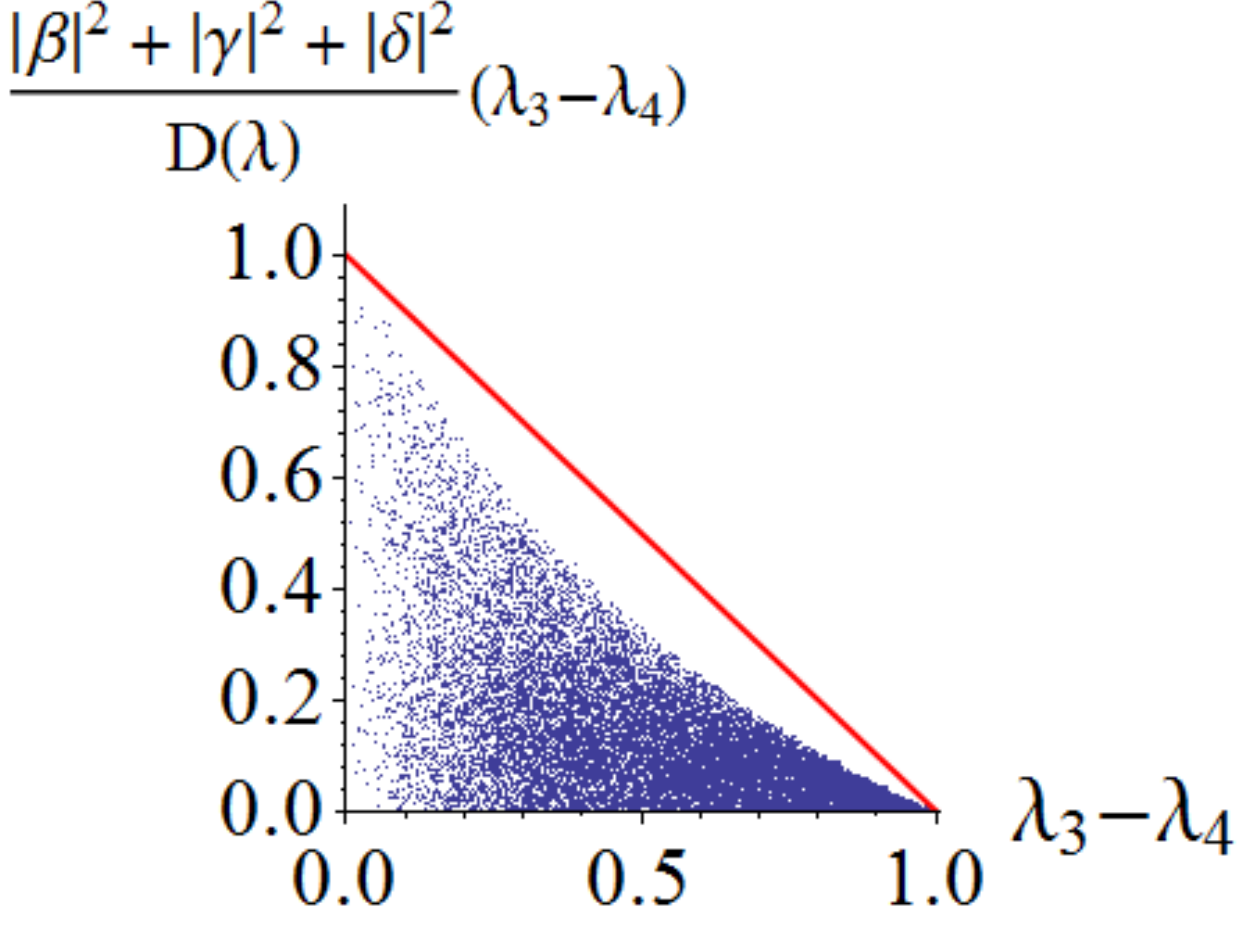}
\captionC{For randomly sampled states $|\Psi_3\rangle \in \wedge^3[\mathcal{H}_1^{(6)}]$ there is shown the ratio of $|\beta|^2+|\gamma|^2+|\delta|^2$ and $D^{(3,6)}(\vec{\lambda})$ as function of $\lambda_3-\lambda_4$ compared to the red curve describing the function $\frac{1}{\lambda_3-\lambda_4}+3$ (on the left). Indeed, for no approximate saturation of $\lambda_3-\lambda_4 \geq 0$ also the three coefficients
$\beta, \gamma$ and $\delta$ are bounded by $D^{(3,6)}(\vec{\lambda})$. From the right, we can infer that the possible divergence of $\frac{|\beta|^2+|\gamma|^2+|\delta|^2}{D^{(3,6)}(\vec{\lambda})}$ is hyperbolic in $\lambda_3-\lambda_4$ and the red straight line shows that one can lower the upper bound in (\ref{unstablebound}) on the scale $\mathcal{O}((\lambda_3-\lambda_4)^0)$.}
\label{fig:Dalt}
\end{figure}

Now, we compare the quantitative prediction of  Theorem \ref{lem:unstablebound} with the numerical results obtained by randomly sampling states $|\Psi_3\rangle \in \wedge^3[\mathcal{H}_1^{(6)}]$ shown in Fig.~\ref{fig:Dalt}. On the left, we plot the relation between $\frac{|\beta|^2+|\gamma|^2+|\delta|^2}{D^{(3,6)}(\vec{\lambda})}$ and $\lambda_3-\lambda_4$. Again, as already shown in Fig.~\ref{fig:36stable}
weights not lying in the $P_{D^{(3,6)}}$-subspace are bounded by $D^{(3,6)}(\vec{\lambda})$ as long as $\lambda_3-\lambda_4 \not \approx 0$. In contrast to the other two weights outside of the $P_{D^{(3,6)}}$-subspace, $|\xi|^2$ and $|\zeta|^2$, the three weights $|\beta|^2, |\gamma|^2$ and $|\delta|^2$ lying on the (green) plane in Fig.~\ref{fig:polytope} are not necessarily quite small when $D^{(3,6)}(\vec{\lambda})\approx 0$ and $(\lambda_3-\lambda_4) \rightarrow 0$ (recall also Ex.~\ref{ex:unstable}). The possible divergence of $\frac{|\beta|^2+|\gamma|^2+|\delta|^2}{D^{(3,6)}(\vec{\lambda})}$ is also compared on the left with the hyperbolic divergence $\frac{1}{\lambda_3-\lambda_4}+3$ (red curve) suggested by Eq.(\ref{unstablebound}). According to the right side of Fig.~\ref{fig:Dalt} and the red straight line shown there,  the bound in Theorem \ref{lem:unstablebound} for $|\beta|^2+|\gamma|^2+|\delta|^2$ can be improved on the basis of the numerical sampling results on the scale $\mathcal{O}((\lambda_3-\lambda_4)^0)$ by reducing it to
\begin{equation}\label{betagammadeltabound}
|\beta|^2+|\gamma|^2+|\delta|^2 \leq \frac{D^{(3,6)}(\vec{\lambda})}{\lambda_3-\lambda_4} - D^{(3,6)}(\vec{\lambda})\,.
\end{equation}
This together with the analytic result (\ref{xizetabound}) in Theorem \ref{lem:xizeta} and the normalization of $|\Psi_3\rangle$ on one hand and Ex.~\ref{ex:Converse} on the other hand finally yields
\begin{equation}\label{alphamunubound}
\frac{D^{(3,6)}(\vec{\lambda})}{2}\leq 1-\big(|\alpha|^2+ |\mu|^2+|\nu|^2\big) \leq \frac{D^{(3,6)}(\vec{\lambda})}{\lambda_3-\lambda_4}\,.
\end{equation}

Eq.~(\ref{alphamunubound}) shows again that Selection Rule (\ref{selrul}) is stable as long as the ordering constraint $\lambda_3\geq \lambda_4$
is not approximately saturated. This also means that an approximate saturation of any other ordering constraint $\lambda_i \geq \lambda_{i+1}$,
$i\neq 3$, cannot lead to instability. But in which sense is the ordering constraint $\lambda_3\geq \lambda_4$ distinguished from all the other ones?
The answer to this question follows immediately from the structure of the corresponding gPC (\ref{d=6b}). The pair $(3,4)$ is the only pair of successive
integers with the property that the corresponding coefficients $\kappa^{(j)}$ (recall Eq.~(\ref{gPC})) in the gPC (\ref{d=6b}) are descending, $0=\kappa^{(3)} > \kappa^{(4)} = -1$. For all the other four pairs $(i,i+1)$, $i\neq 3$, we have $\kappa^{(i)}\leq \kappa^{(i+1)}$. This insight suggests an extension of Selection Rule (\ref{selrul}) for all $N$ and all $d$ :

\begin{conjec}[Extension of Selection Rule (\ref{selrul})]\label{sel2}
Consider an $N$-fermion state $|\Psi_N\rangle$ with NO $\{|i\rangle\}_{i=1}^d$ and NON $\vec{\lambda}$ with one degeneracy, $\lambda_i=\lambda_{i+1}$, and saturating a gPC $D(\cdot)\geq 0$. Using the self-consistent expansion (\ref{Psic}), $|\Psi_N\rangle = \sum_{\textbf{i}}\,c_{\textbf{i}}\,|\textbf{i}\,\rangle$ we have (recall Eq.~(\ref{Dop}))
\begin{itemize}
\item if $\kappa^{(i)}\leq \kappa^{(i+1)}$:
\begin{equation}
\hat{D}_{\Psi_N} |\textbf{i}\,\rangle \neq 0 \,\,\qquad \,\,\Rightarrow \qquad c_{\textbf{i}} = 0
\end{equation}
\item if $\kappa^{(i)} > \kappa^{(i+1)}$:
\begin{equation}\label{kappabad}
\begin{array}{r}
\hat{D}_{\Psi_N}|\textbf{i}\,\rangle \neq0 \\
\wedge \,\,\, \hat{D}_{\Psi_N} \hat{\pi}_{i,i+1}|\textbf{i}\,\rangle \neq 0
\end{array}\biggr\}
\,\,\,\, \Rightarrow \,\,\, c_{\textbf{i}} = 0\,,
\end{equation}
\end{itemize}
Here, $ \hat{\pi}_{i,i+1}$ is the operator which swaps the NO $|i\rangle$ and $|i+1\rangle$ which may appear in a Slater determinant
$|\textbf{i}\,\rangle$.
\end{conjec}
In the next subsection we provide further numerical evidence for the validity of this conjecture.


\subsection{Larger settings}\label{sec:largersettings}
We investigate whether the results found in Sec.~\ref{sec:BDsetting} for the Borland-Dennis setting $\wedge^3[\mathcal{H}_1^{(6)}]$ also hold
in larger settings. Since the estimation techniques we used to prove Theorem \ref{lem:xizeta} and Theorem \ref{lem:unstablebound} cannot easily be generalized to larger settings we resort to numerical methods. This in particular also reflects the fact explained at the beginning of Sec.~\ref{sec:BDsetting} that larger settings are less relevant for physical applications due to the unmanageable and too specific structure of their polytopes.

By sampling quantum states in $\wedge^3[\mathcal{H}_1^{(7)}]$ and $\wedge^3[\mathcal{H}_1^{(8)}]$ we can explore the possible stability of Selection Rule (\ref{selrul}) for the next larger settings.

\subsubsection{Three fermions and seven dimensions}\label{sec:setting37}
For the setting $\wedge^3[\mathcal{H}_1^{(7)}]$ the gPC for the NON $\vec{\lambda}\equiv (\lambda_1,\ldots,\lambda_7)$
are given by \cite{Borl1972,Kly3}
\begin{eqnarray}\label{gPC37}
D_1^{(3,7)}(\vec{\lambda}) &=& 2-(\lambda_1+\lambda_2+\lambda_5+\lambda_6)\geq 0 \nonumber \\
D_2^{(3,7)}(\vec{\lambda}) &=& 2-(\lambda_1+\lambda_3+\lambda_4+\lambda_6)\geq 0 \nonumber \\
D_3^{(3,7)}(\vec{\lambda}) &=& 2-(\lambda_2+\lambda_3+\lambda_4+\lambda_5)\geq 0 \nonumber \\
D_4^{(3,7)}(\vec{\lambda}) &=& 2-(\lambda_1+\lambda_2+\lambda_4+\lambda_7)\geq 0\,.
\end{eqnarray}
\begin{figure}[]
\centering
\includegraphics[width=0.22\textwidth]{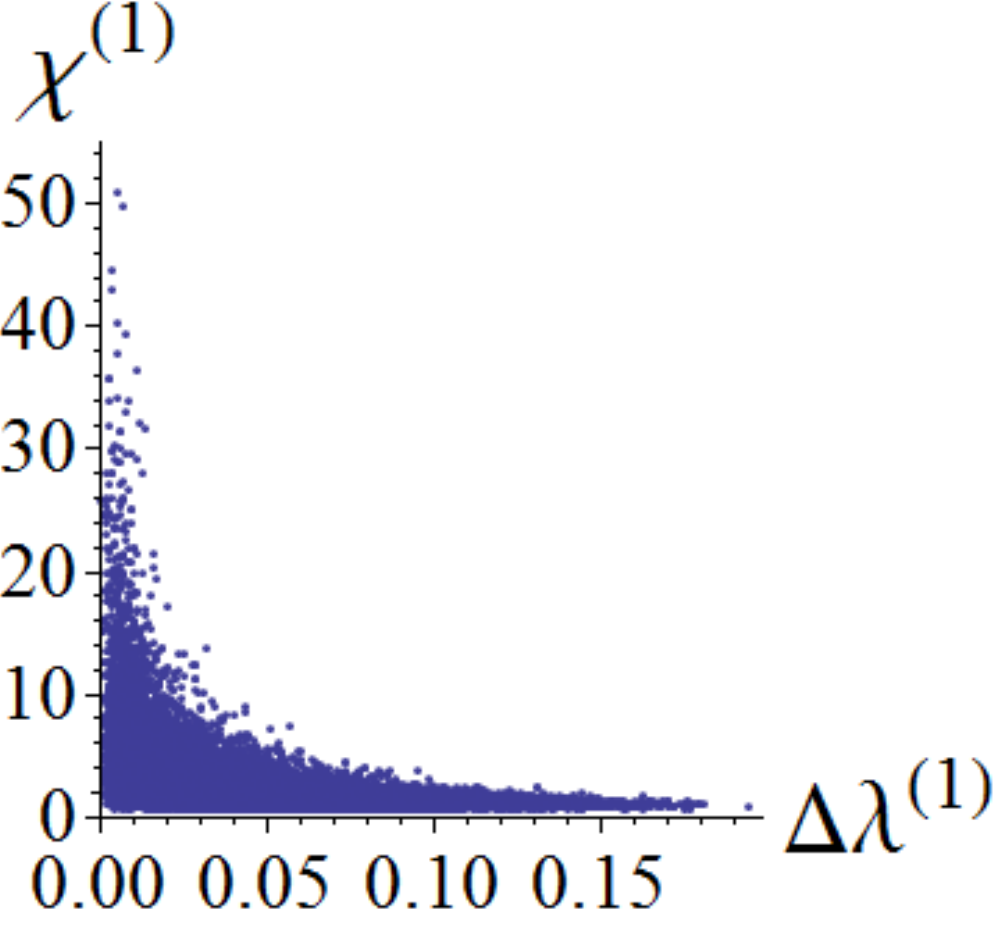}\hspace{0.2cm}
\includegraphics[width=0.22\textwidth]{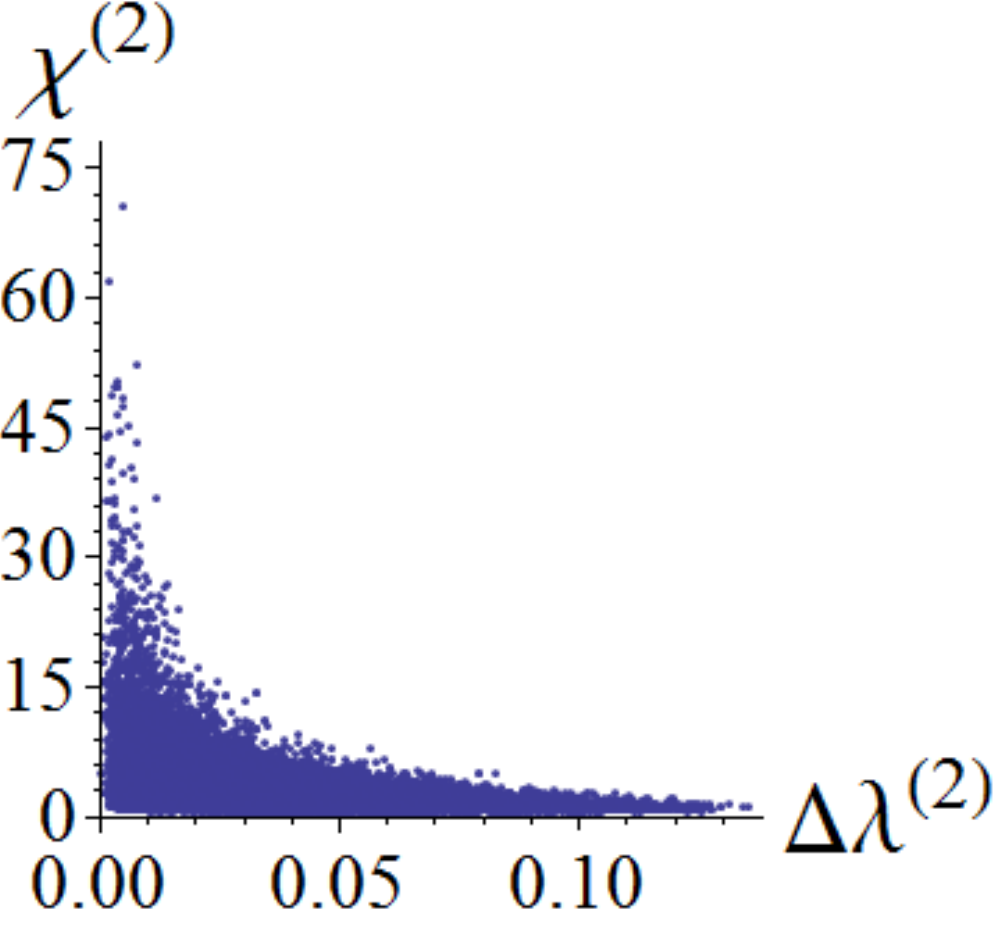}
\includegraphics[width=0.22\textwidth]{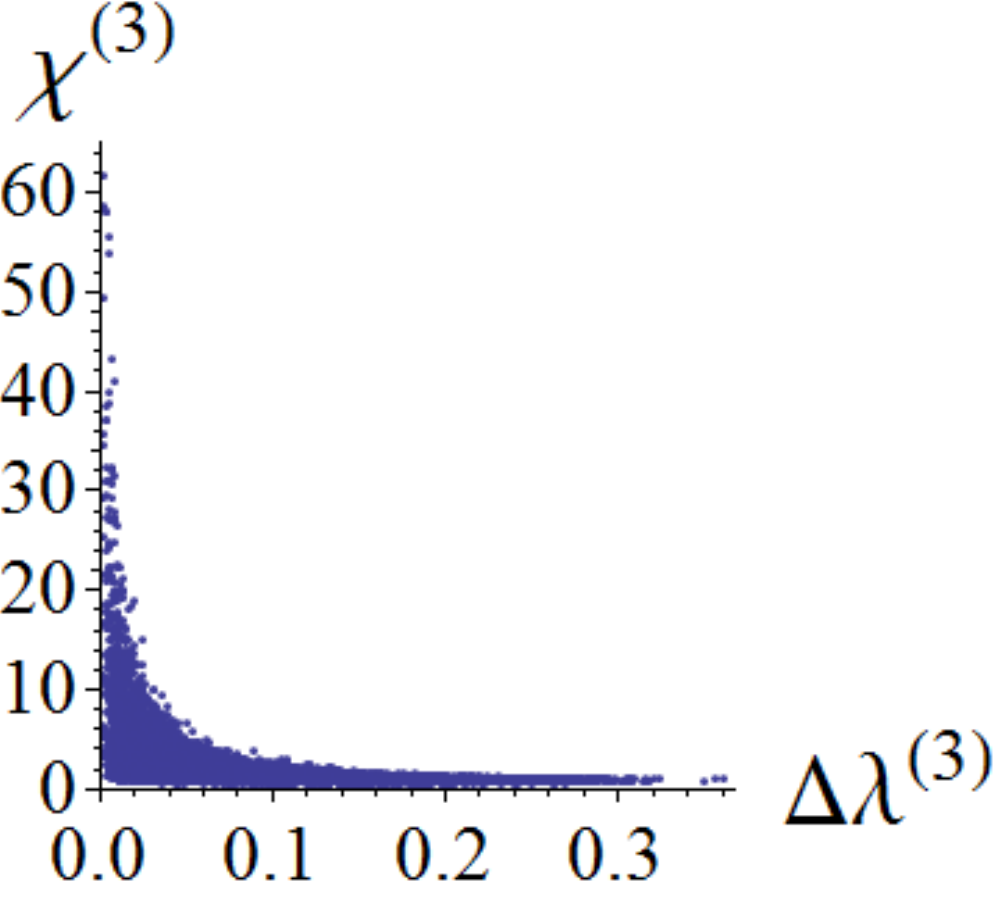}\hspace{0.2cm}
\includegraphics[width=0.22\textwidth]{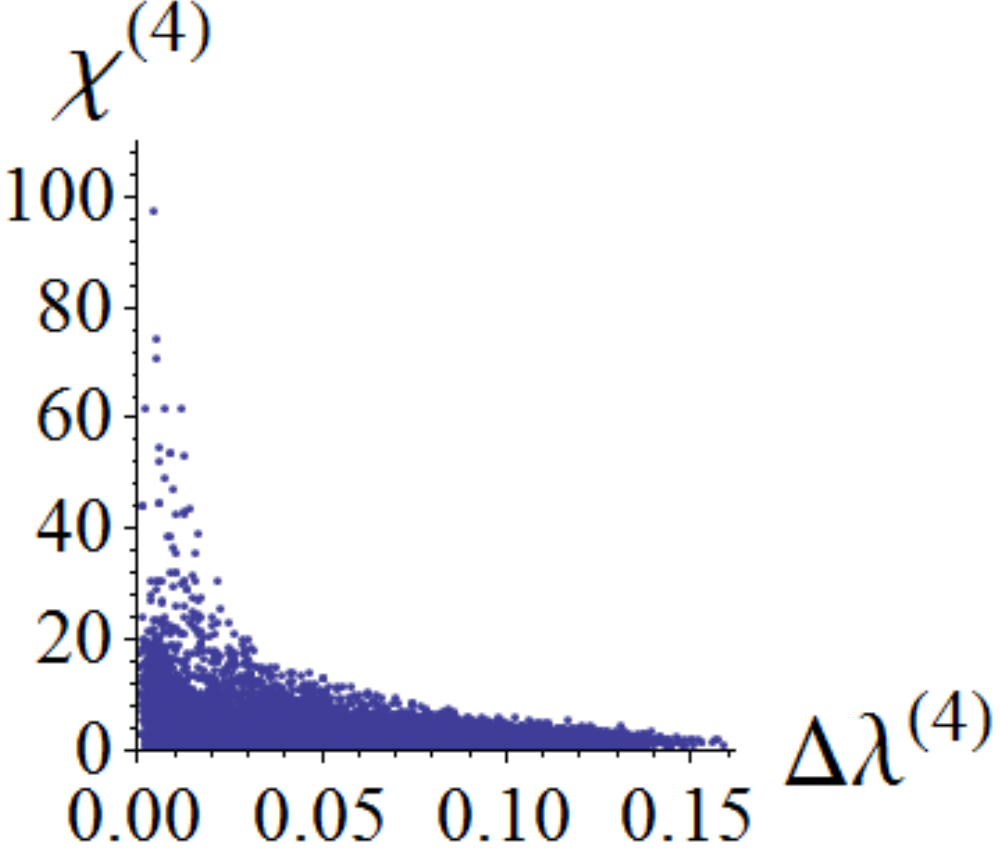}
\captionC{For each of the four gPC (\ref{gPC37}) of the setting $\wedge^3[\mathcal{H}_1^{(7)}]$ the stability of Selection Rule (\ref{selrul}) is explored. For randomly sampled states exhibiting quasipinning of strength $D_j^{(3,7)}(\vec{\lambda})\leq 0.01$ we study the ratio $\chi^{(j)}$ of the weight $1-\|P_{D_j^{(3,7)}}\Psi_3\|_{L^2}^2$ outside the subspace corresponding to pinning and $D_j^{(3,7)}(\vec{\lambda})$. The behavior of the indicators $\chi^{(j)}$ show that Selection Rule (\ref{selrul}) is stable unless some specific ordering constraints $\lambda_i-\lambda_{i+1}\geq 0$ (see also (\ref{DeltaNON})) are  approximately saturated.}
\label{fig:37Da}
\end{figure}
\noindent We randomly sampled $2\times10^9$ states $|\Psi_3\rangle \in \wedge^3[\mathcal{H}_1^{(7)}]$ and study their structure whenever their NON are quasipinned by one of the four gPC (\ref{gPC}). By choosing $D_j^{(3,7)}(\vec{\lambda}) \leq 0.01$ as criterion for quasipinning we found about $2\times10^6$ quasipinned states. For these $|\Psi_3\rangle$ we calculate the corresponding weights $\|P_{D_j^{(3,7)}}\Psi_3\|_{L^2}^2$
lying in the $P_{D_j^{(3,7)}}$-subspaces (recall Def.~\ref{def:projD=0}). We also calculate
the corresponding indicator of stability for Selection Rule (\ref{selrul}),

\begin{equation}\label{indicator}
\chi^{(j)} \equiv \frac{1-\|P_{D_j^{(3,7)}}\Psi_3\|_{L^2}^2}{D_j^{(3,7)}(\vec{\lambda})}\,.
\end{equation}

\begin{figure}[]
\centering
\includegraphics[width=0.22\textwidth]{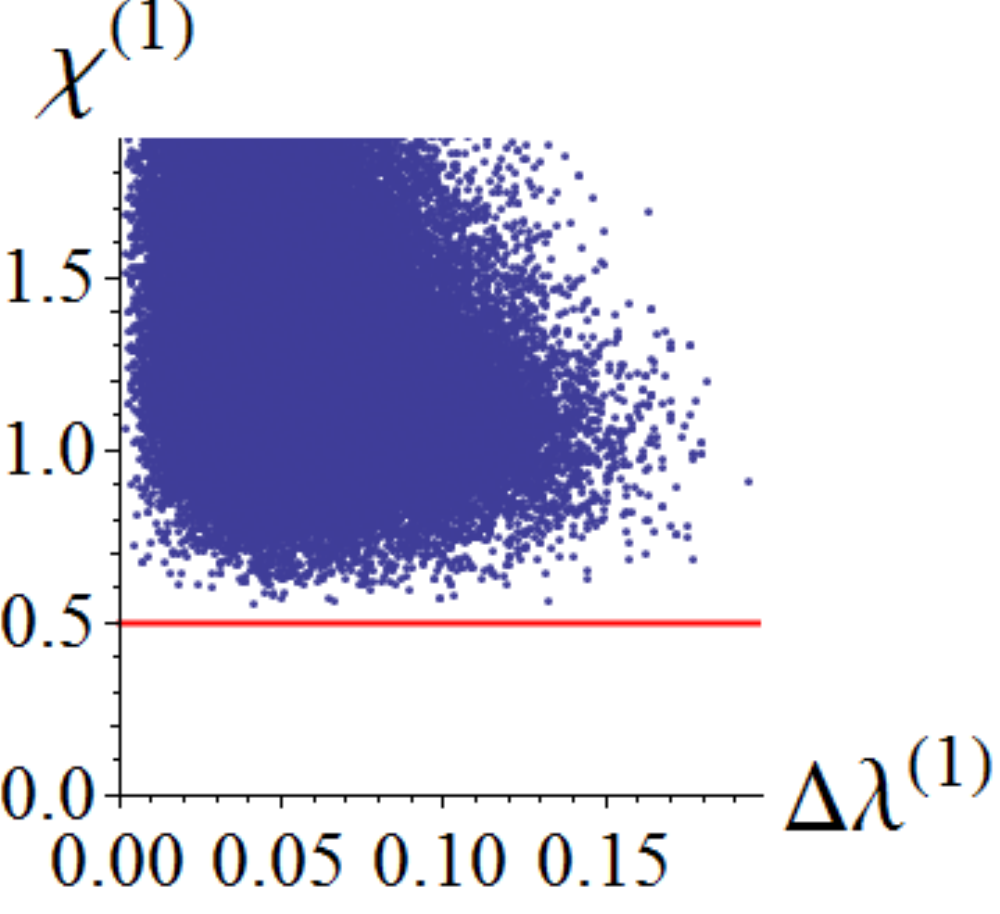}\hspace{0.2cm}
\includegraphics[width=0.22\textwidth]{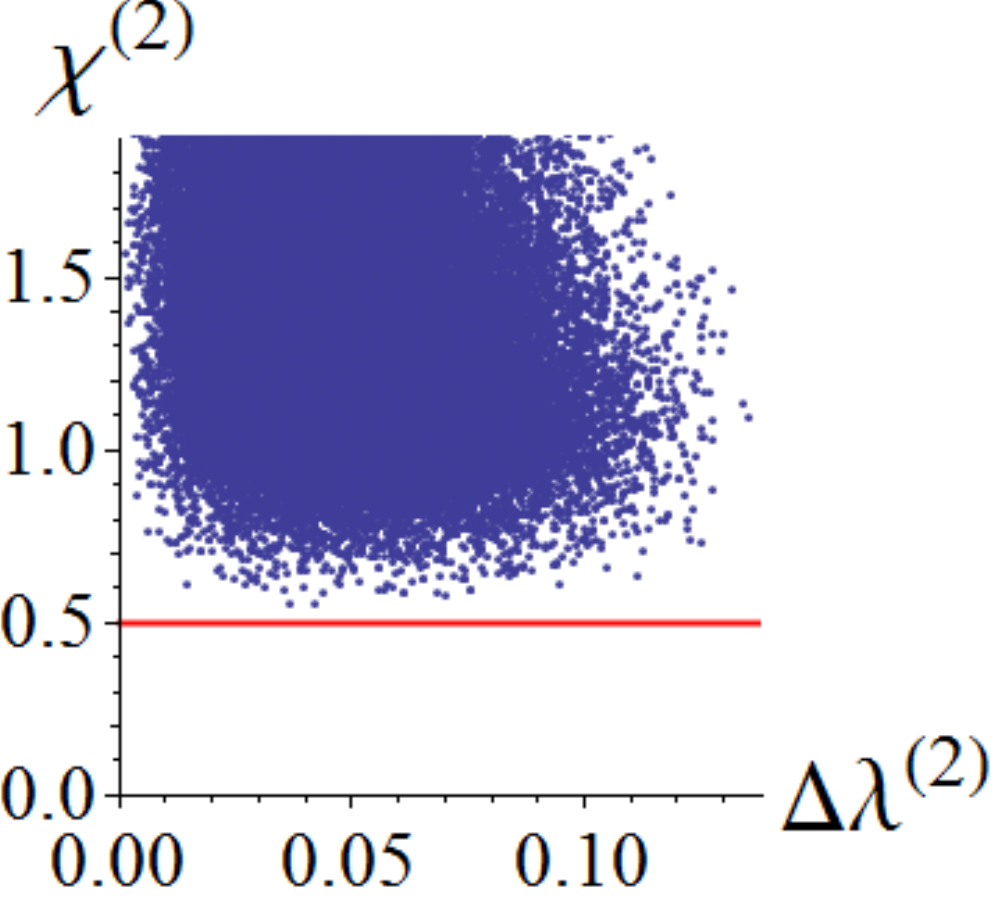}
\includegraphics[width=0.22\textwidth]{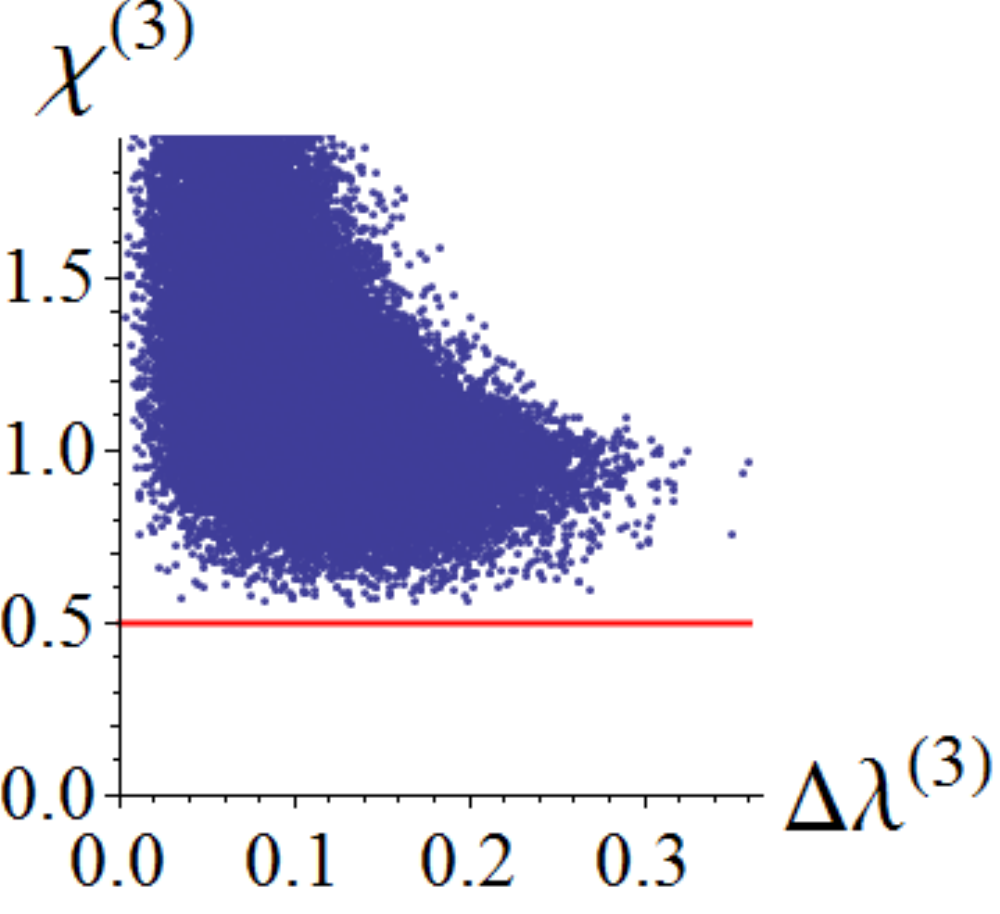}\hspace{0.2cm}
\includegraphics[width=0.22\textwidth]{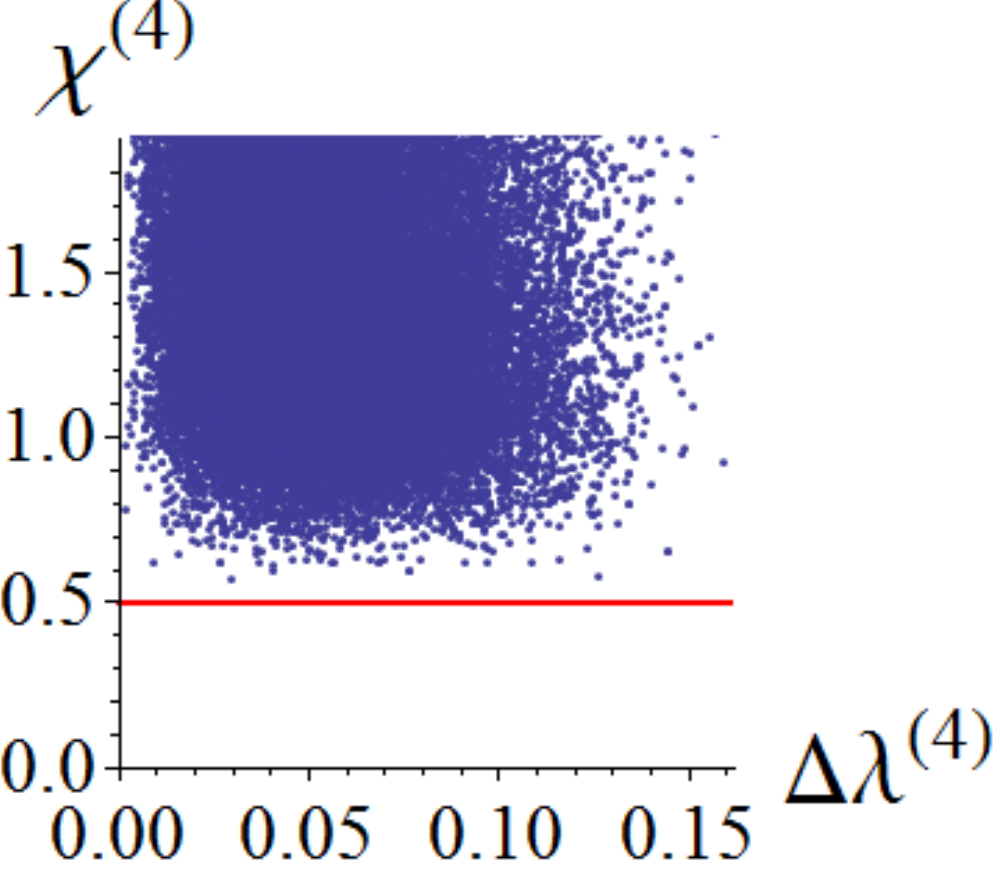}
\captionC{Same figure as Fig.~\ref{fig:37Da}. However, the smaller range for the indicators $\chi^{(i)}$ allows to conclude a lower bound (red straight line) on the indicators $\chi^{(j)}$ quantifying that finite $D_j^{(3,7)}(\vec{\lambda})$ (i.e.~quasipinning) implies a finite weight of the corresponding quantum state $|\Psi_3\rangle$ outside the pinning subspace described by $P_{D_j^{(3,7)}}$ (recall Def.~\ref{def:projD=0}).}
\label{fig:37Db}
\end{figure}

The selection rule is stable in the vicinity of the polytope boundary where $\chi_j \leq \mathcal{O}(1)$. In Fig.~\ref{fig:37Da} the indicators $\chi^{(j)}$ are shown for all four gPC (\ref{gPC37}) as a function of the following eigenvalue distances $\Delta \lambda^{(j)}$
\begin{eqnarray}\label{DeltaNON}
\Delta \lambda^{(1)} \equiv \lambda_4-\lambda_5,\, \Delta \lambda^{(2)} \equiv \mbox{min}(\lambda_2-\lambda_3,\lambda_5-\lambda_6)&& \nonumber \\
\Delta \lambda^{(3)} \equiv \lambda_1-\lambda_2,\, \Delta \lambda^{(4)} \equiv \mbox{min}(\lambda_3-\lambda_4,\lambda_6-\lambda_7)&&\,.
\end{eqnarray}

That we consider exactly those differences of NON is suggested by Conjecture \ref{sel2} and the concrete form of the coefficients $\kappa_j^{(i)}$ in Eq.(\ref{gPC37}). In Fig.~\ref{fig:37Da} we clearly see that the indicators $\chi^{(j)}$ are bounded by $\mathcal{O}(1)$ as long as $\Delta \lambda^{(j)} \not \approx 0$. Again, as already seen for the Borland-Dennis setting, Selection Rule (\ref{selrul}) is stable as long as none of the ordering constraints is approximately saturated. In Fig.~\ref{fig:37Db} the sampled results are presented again, but for a smaller range which shows that the lower bound on the stability of Selection Rule (\ref{selrul}) provided by Theorem \ref{stat:ConverseStab} and Ex.~\ref{ex:Converse} is tight for the setting $\wedge^3[\mathcal{H}_1^{(7)}]$.

\begin{figure}[]
\centering
\includegraphics[width=0.22\textwidth]{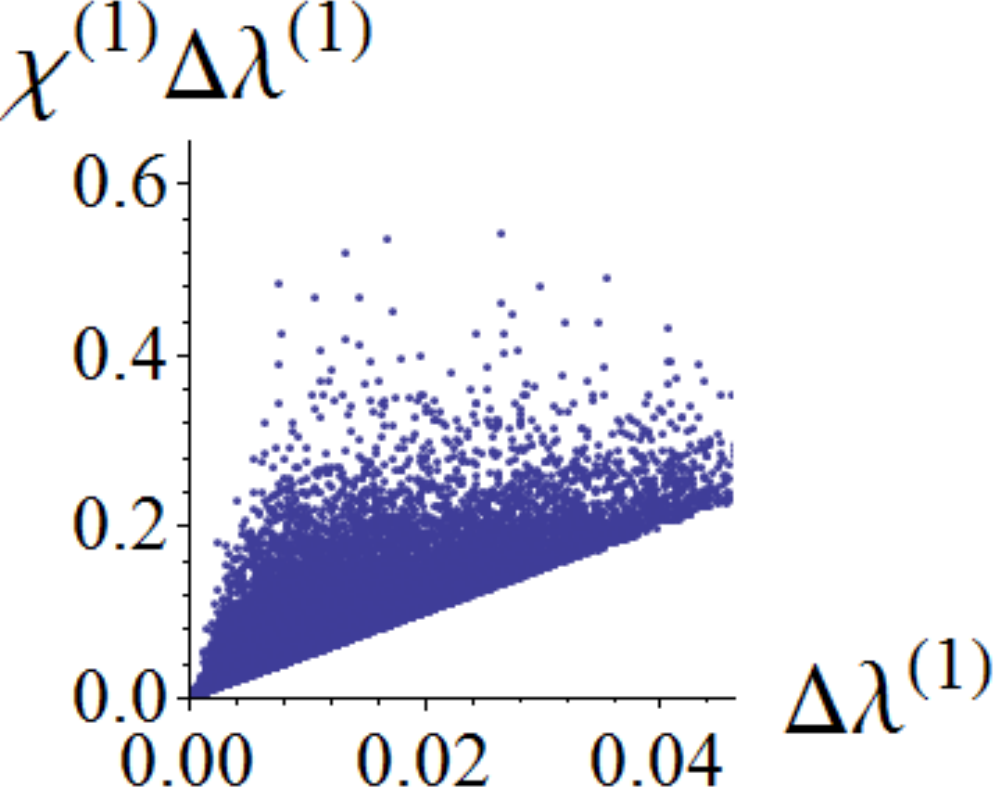}\hspace{0.2cm}
\includegraphics[width=0.22\textwidth]{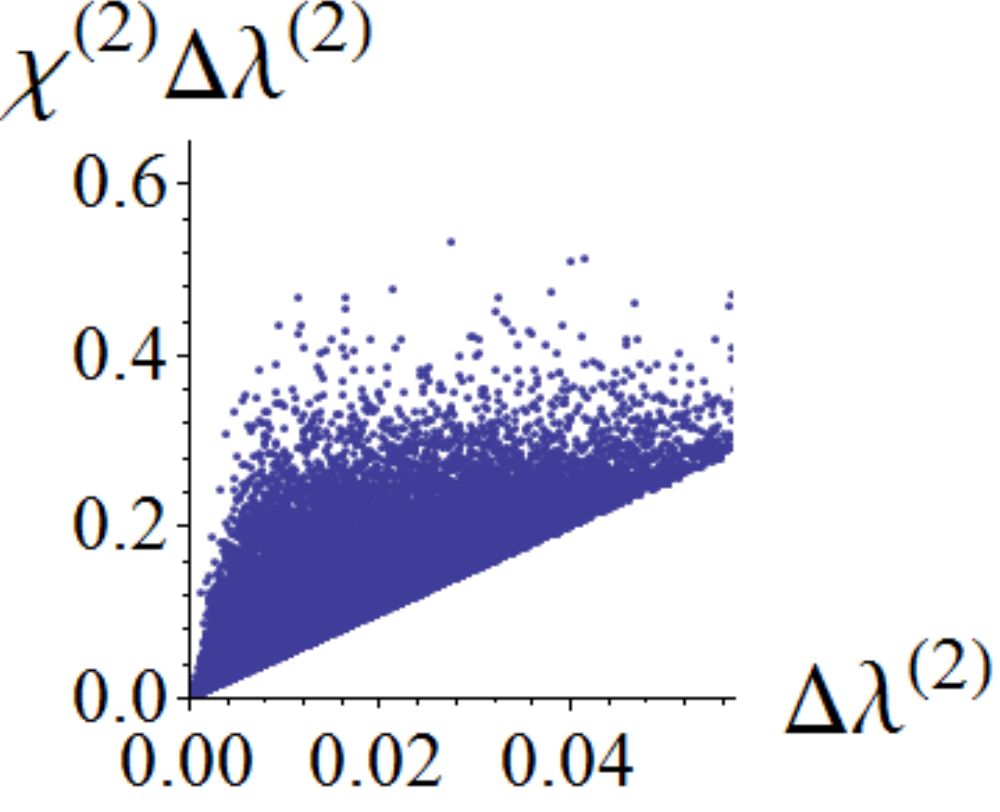}
\includegraphics[width=0.22\textwidth]{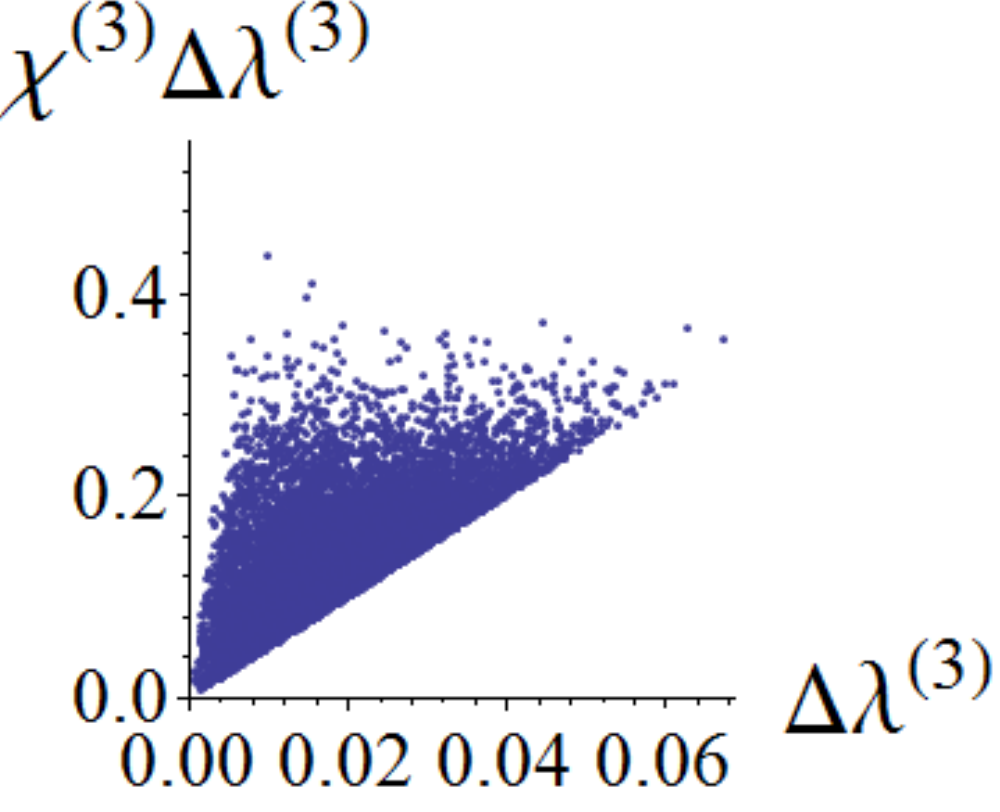}\hspace{0.2cm}
\includegraphics[width=0.22\textwidth]{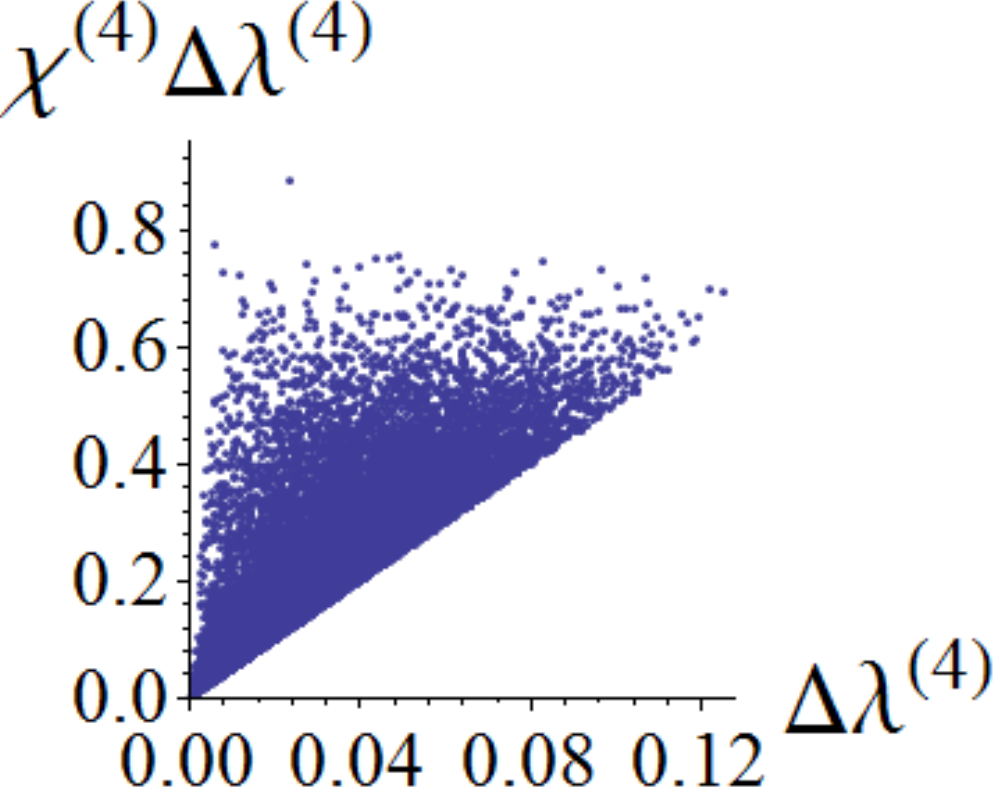}
\captionC{The divergence at $\Delta\lambda^{(i)}=0$ of the stability indicators $\chi^{(i)}$ shown in Fig.~\ref{fig:37Da} is investigated for the sampled states. Since $\chi^{(i)} \Delta\lambda^{(i)}$ is bounded from above by $\mathcal{O}(1)$ this divergence is hyperbolic in $\Delta\lambda^{(i)}$. }
\label{fig:37Dc}
\end{figure}

The divergence of the indicators $\chi^{(j)}$ is demonstrated in Fig.~\ref{fig:37Dc}. There, we find strong numerical evidence that the possible divergence of $\chi^{(j)}$ at $\Delta \lambda^{(j)} = 0$ is bounded by $\frac{1}{\Delta \lambda^{(j)}}$.
All these numerical insights imply the following bounds on stability of Selection Rule (\ref{selrul}) (recall Def.~\ref{def:projD=0}, \mbox{Theorem \ref{stat:ConverseStab}} and Eq.~(\ref{DeltaNON}))
\begin{equation}\label{stabilityall37}
\frac{D_j^{(3,7)}(\vec{\lambda})}{2}\leq 1-\|P_{D_j^{(3,7)}} \Psi_3\|_{L^2}^2 \leq \frac{D_j^{(3,7)}(\vec{\lambda})}{\Delta \lambda^{(j)}}\,,
\end{equation}
where $j=1,2,3,4$.

Finally, we verify the validity of Conjecture \ref{sel2}.
For the case $\kappa^{(i)}>\kappa^{(i+1)}$ we consider exemplary the gPC $D_1^{(3,7)}(\cdot)\geq 0$ with ordering constraint
$\Delta \lambda^{(1)}\equiv \lambda_4-\lambda_5 \geq 0$ and gPC $D_3^{(3,7)}(\cdot)\geq 0$ with ordering constraint $\Delta \lambda^{(3)}\equiv \lambda_1-\lambda_2 \geq 0$ (recall Eq.~(\ref{gPC37})).
Validity of Conjecture \ref{sel2} implies that for $\Delta \lambda^{(j)}\approx 0$ the Slater determinants neither lying in the $P_{D_j^{(3,7)}}$- nor in the $P_{\pi_{i,i+1}D_j^{(3,7)}}$-subspace (where $i=4$ for $j=1$ and $i=1$ for $j=3$) have weights which are bounded by $D_j^{(3,7)}(\vec{\lambda})$. Let us denote the sum of the squares of these weights lying in none of these two subspaces by $W_1$ and $W_3$
corresponding to the gPC $D_1^{(3,7)}(\cdot)\geq 0$ and $D_3^{(3,7)}(\cdot)\geq 0$, respectively.
\begin{figure}[]
\vspace{0.4cm}
\centering
\includegraphics[width=0.22\textwidth]{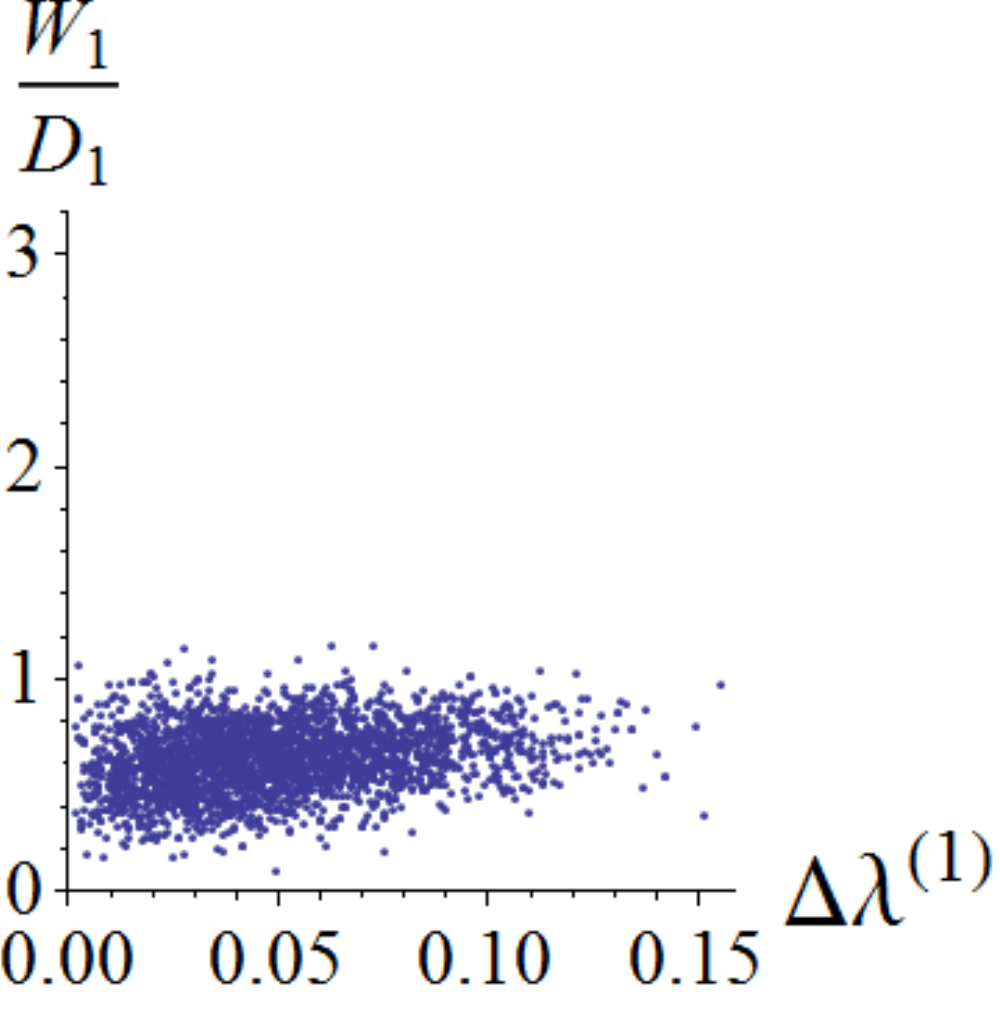}\hspace{0.2cm}
\includegraphics[width=0.22\textwidth]{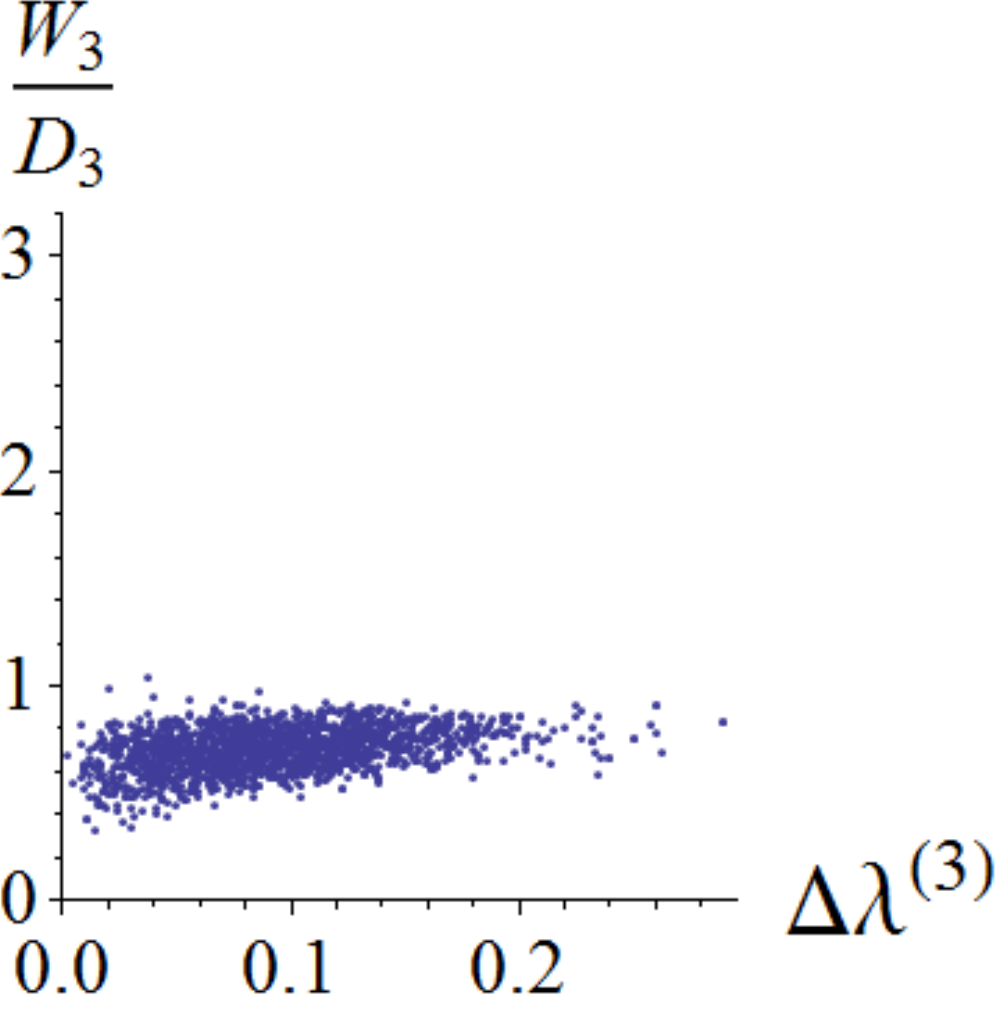}
\captionC{Stability of the statement made by Conjecture \ref{sel2} for the case $\kappa^{(i)} > \kappa^{(i+1)}$ is numerically found for the setting $\wedge^3[\mathcal{H}_1^{(7)}]$
by exemplarily studying the first (left) and third (right) gPC in Eq.~(\ref{gPC37}). $W_j$ is the weight of all Slater determinants which should be suppressed whenever $D_j(\vec{\lambda})$ is small. $\Delta \lambda^{(j)}$ is given by Eq.~(\ref{DeltaNON}).}
\label{fig:37Dd}
\end{figure}
In the corresponding analysis, to minimize inaccuracies due to the finiteness of $D_j^{(3,7)}(\vec{\lambda})$ we lowered the threshold for quasipinning from originally $D_j^{(3,7)} \leq 0.01$ to $D_j^{(3,7)} \leq 0.005$. Moreover, to deal only with the case of one quasisaturated ordering constraint (as required by Conjecture \ref{sel2}) we need to exclude all sampled results with an additional quasisaturation of another ordering constraint. Actually, it was sufficient to just restrict the analysis of $j=1$ to $\lambda_5-\lambda_6 \geq 0.05$ and the analysis of $j=3$ to $\lambda_2-\lambda_3 \geq 0.05$. The corresponding results are shown in Fig.~\ref{fig:37Dd}. Since the ratios $\frac{W_j}{D_j^{(3,7)}}$, $j=1,3$ are both bounded by $\mathcal{O}(1)$ universal in $\Delta \lambda^{(j)}$ Conjecture \ref{sel2} is valid for the case $\kappa^{(i)} > \kappa^{(i+1)}$. That Conjecture \ref{sel2} is also valid for the case $\kappa^{(i)} \leq \kappa^{(i+1)}$ has already been seen in Fig.~\ref{fig:37Da} since there only quasisaturation of an ordering constraint $\lambda_k \geq \lambda_{k+1}$ with $\kappa^{(k)} > \kappa^{(k+1)}$ could have caused violation of Selection Rule (\ref{selrul})  which makes the same predictions as Conjecture \ref{sel2} for  $\kappa^{(i)} \leq \kappa^{(i+1)}$.

\subsubsection{Three fermions and eight dimensions}\label{sec:setting38}
For the setting $\wedge^3[\mathcal{H}_1^{(8)}]$ there are 31 gPC for the NON $\vec{\lambda}\equiv (\lambda_1,\ldots,\lambda_8)$
presented in \cite{Kly3}. Four of them are identical to the four gPC (\ref{gPC37}) of the setting $\wedge^3[\mathcal{H}_1^{(7)}]$.
A fifth one, which will play a role for the numerical analysis takes the form
\begin{equation}\label{gPC38specific}
D_5^{(3,8)}(\vec{\lambda}) = 1-(\lambda_1+\lambda_8)\geq 0\,.
\end{equation}
\noindent We randomly sampled $12.8\times 10^9$ states $|\Psi_3\rangle \in \wedge^3[\mathcal{H}_1^{(8)}]$ and studied their structure whenever their NON are quasipinned by one of the corresponding 31 gPC \cite{Kly3}. Although we chose here a much larger threshold for the definition of quasipinning, $D_j^{(3,8)}(\vec{\lambda}) \leq 0.05$, we find only about $25800$ quasipinned states. For these $|\Psi_3\rangle$ we calculate again the corresponding weights $\|P_{D_j^{(3,8)}}\Psi_3\|_{L^2}^2$ lying in the $P_{D_j^{(3,8)}}$-subspace (recall Def.~\ref{def:projD=0}) and the corresponding indicators $\chi^{(j)}$(\ref{indicator})
for stability of Selection Rule (\ref{selrul}).

\begin{center}
\begin{figure}[h]
\centering
\includegraphics[width=0.18\textwidth]{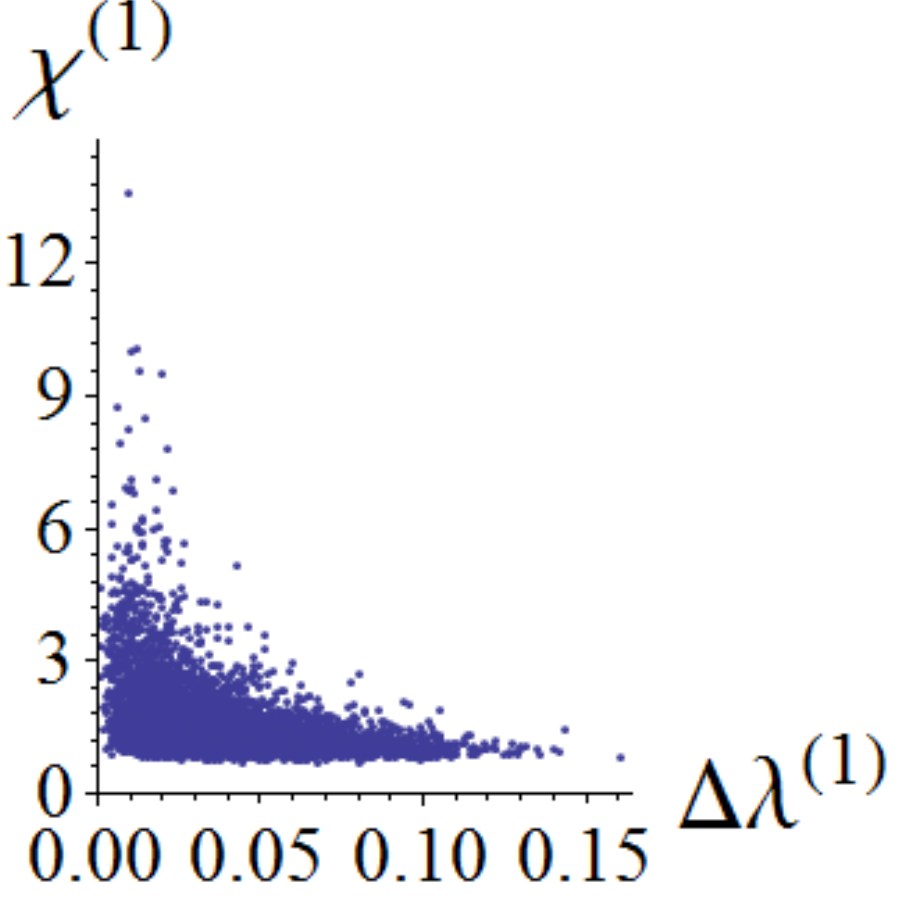}\hspace{0.1cm}
\includegraphics[width=0.18\textwidth]{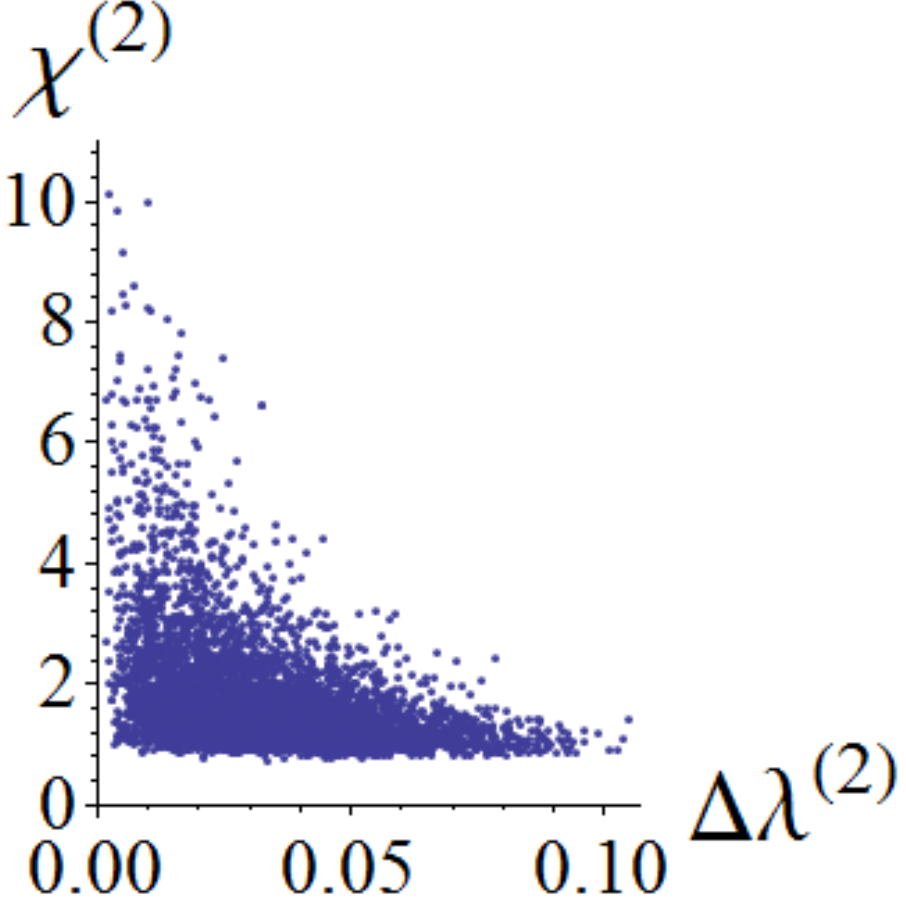}\hspace{0.1cm}
\includegraphics[width=0.18\textwidth]{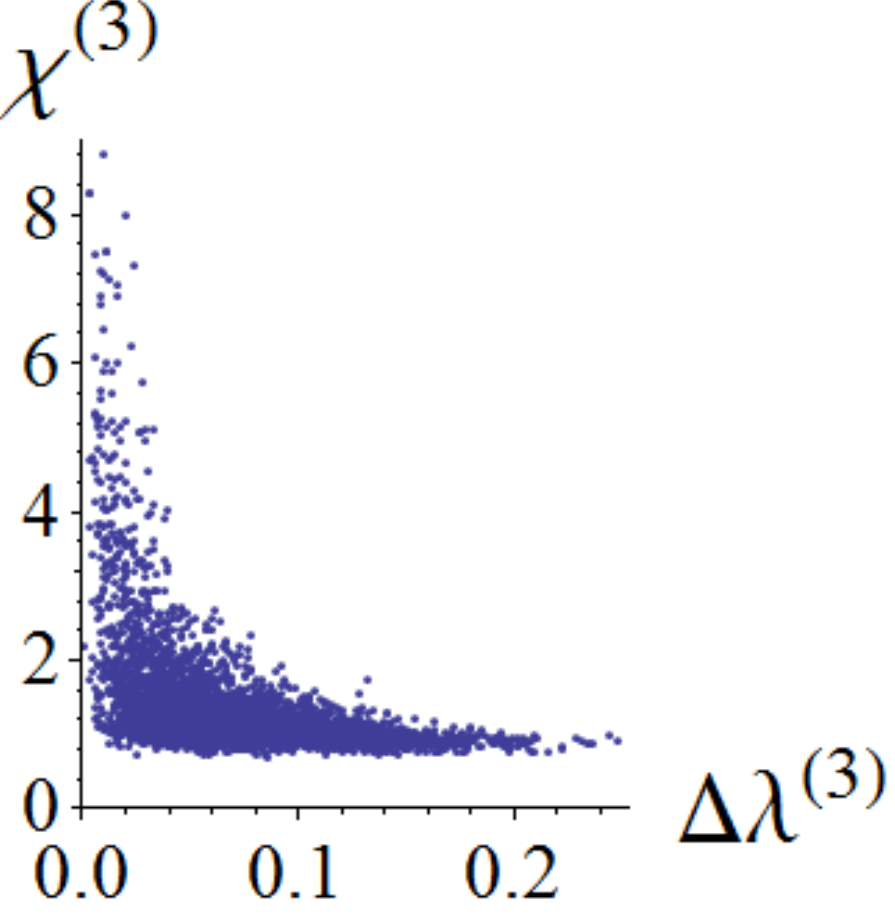}\hspace{0.1cm}
\includegraphics[width=0.18\textwidth]{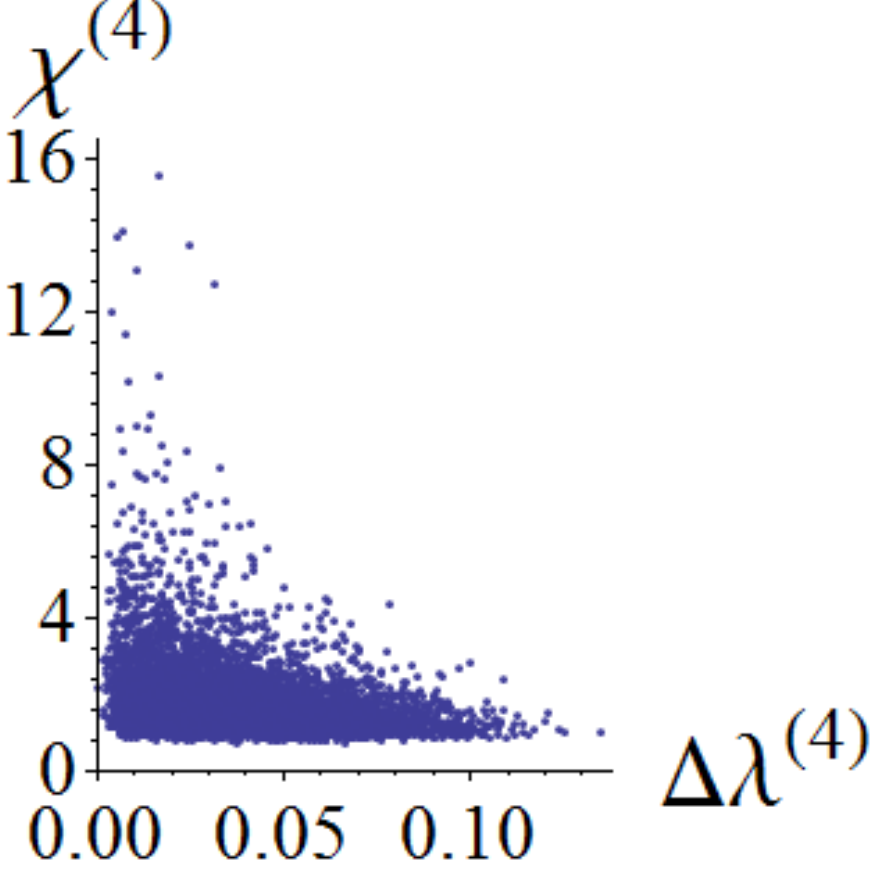}\hspace{0.1cm}
\includegraphics[width=0.18\textwidth]{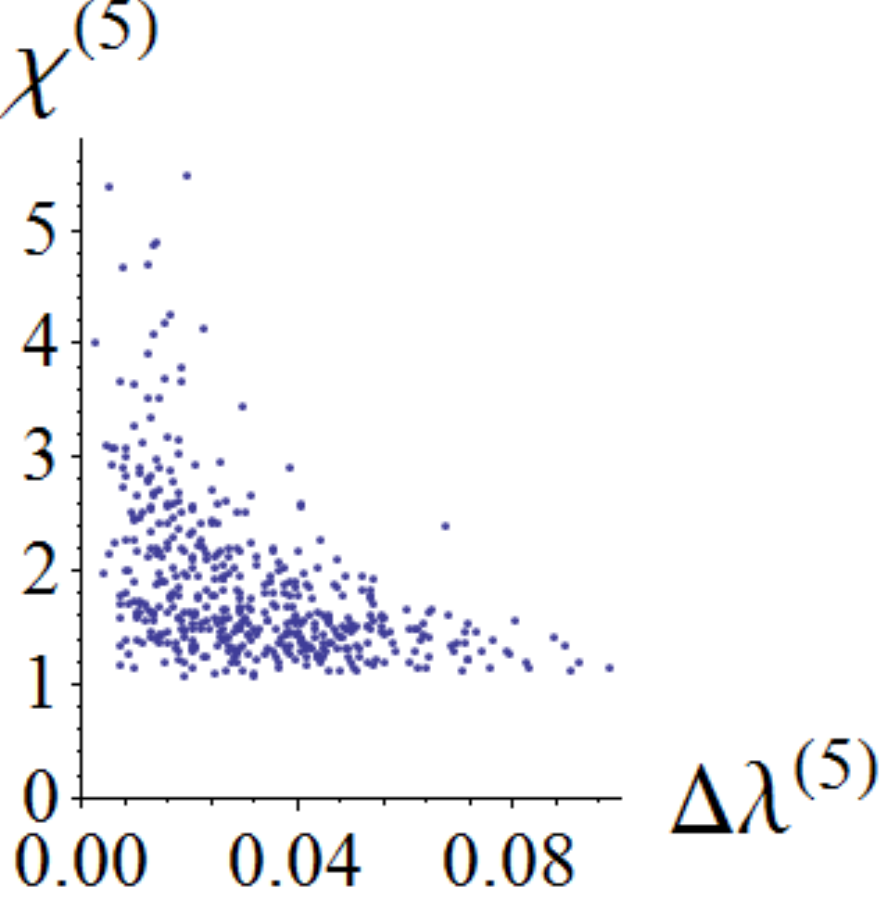}
\captionC{For five gPC of the setting $\wedge^3[\mathcal{H}_1^{(8)}]$ the stability of Selection Rule (\ref{selrul}) is explored. For randomly sampled states exhibiting quasipinning of strength $D_j^{(3,8)}(\vec{\lambda})\leq 0.05$ we study the ratio $\chi^{(j)}$ of the weight $1-\|P_{D_j^{(3,8)}}\Psi_3\|_{L^2}^2$ outside the subspace corresponding to pinning and $D_j^{(3,8)}(\vec{\lambda})$. The behavior of the indicators $\chi^{(j)}$ shows that Selection Rule (\ref{selrul}) is stable unless some specific ordering constraints $\lambda_i-\lambda_{i+1}\geq 0$ (see also (\ref{DeltaNON})) are  approximately saturated.}
\label{fig:38Da}
\end{figure}
\end{center}

The results are shown in Fig.~\ref{fig:38Da} for $D_j^{(3,8)}, j=1,...,5$ where the differences $\Delta \lambda^{(j)}$ for $j=1,2,3,4$ are given by Eq.~(\ref{DeltaNON}) and $\Delta \lambda^{(5)} \equiv \lambda_7-\lambda_8$. The indicators $\chi^{(j)}$ are again bounded from above by $\mathcal{O}(1)$ as long as $\Delta \lambda^{(j)} \not \approx 0$. Hence,
Selection Rule (\ref{selrul}) is stable as long as no ordering constraint $\lambda_{i}\geq \lambda_{i+1}$
with $\kappa_j^{(i)}>\kappa_{j}^{(i+1)}$ is approximately saturated. Moreover, it turns out that the possible divergence of $\chi^{(j)}$ at $\Delta \lambda^{(j)}=0$ is hyperbolic again.

From the sampled results for gPC (\ref{gPC38specific}) and the fact that we also relaxed the threshold for quasipinning to $D(\vec{\lambda})\leq 0.05$ it can be seen that for the present setting $\wedge^3[\mathcal{H}_1^{(8)}]$ randomly sampled states are typically not quasipinned. In particular there was even no single quasipinned state for most of the other $26$ gPC. Consequently it seems that quantitative bounds for the stability Selection Rule (\ref{selrul}) and validity of Conjecture \ref{sel2} can not be found for larger settings using the present numerical approach.

Notice also that this remark on randomly sampled quantum states does not imply at all that ground states of few-fermion systems with larger $1$-particle Hilbert spaces do never exhibit quasipinning. As already explained  below Remark \ref{rem:PinArtificial} ground states are quite unique since they are the minimizer of the energy expectation value and therefore could have quasipinned NON.

\onecolumngrid
\twocolumngrid

\section{Summary and Conclusion}\label{sec:concl}
The antisymmetry of $N$-fermion quantum states under particle exchange does not only imply Pauli's famous exclusion principle but leads to even stronger restrictions of fermionic occupation numbers. These so-called generalized Pauli constraints (gPC) form a polytope of possible natural occupation numbers (NON), the eigenvalues of the $1$-particle reduced density operator. The gPC may be relevant for concrete physical applications whenever the NON of a given quantum state are pinned to the polytope's boundary. In such a case the physical system is limited in the sense that the (vector of) NON cannot evolve in any arbitrary direction under some given time evolution. Another important aspect was mentioned by Klyachko \cite{Kly1} at least for the case of non-degenerate NON: Pinning of NON implies that the corresponding $N$-fermion quantum state $|\Psi_N\rangle$ has a very specific and significantly simpler structure. By expanding $|\Psi_N\rangle$ w.r.t.~Slater determinants built up from its own natural orbitals only a few very specific Slater determinants are allowed to contribute. This selection rule of Slater determinants generalizes the well-known result that NON $\vec{\lambda}=(1,\ldots,1,0,\ldots)$ can only arise from a single Slater determinant $|\Psi_N\rangle = |1,2,\ldots,N\rangle$.

Although pinning would have remarkable physical consequences we argue that its occurrence for concrete systems is very unlikely due to its unrealistic rigorous structural implications for the corresponding $|\Psi_N\rangle$ (see Remark \ref{rem:PinArtificial}). Recently, analytic evidence was found \cite{CS2013} that NON of fermionic ground states may lie close but not exactly on the polytope boundary (i.e.~they are quasipinned).

Therefore, the major motivation of the present work was to investigate how far quasipinning, in analogy to pinning, determines the structure of the corresponding $N$-fermion state, i.e.~to study the stability of Selection Rule (\ref{selrul}). We have used analytical and numerical methods.

To quantify stability of Selection Rule (\ref{selrul}) we have determined lower and upper bounds for various weights of Slater determinants contributing to an $N$-fermion quantum state $|\Psi_N\rangle$. The projection  $\|P_{D}\Psi_N\|_{L^2}$
of  $|\Psi_N\rangle$ onto the zero-subspace of the operator $\hat{D}_{\Psi_N}$ corresponding to a gPC  $D(\vec{\lambda})$
plays a particular role. For \emph{arbitrary} $N$ and $d$ we have proven that $(1- \|P_{D}\Psi_N\|_{L^2}^2)$ is bounded from below by $\frac{D(\vec{\lambda})}{\|\hat{D}_{\Psi_N}\|_{\small{\mbox{op}}}}$ (Theorem \ref{stat:ConverseStab}).
This means that the weight of all Slater determinants contributing to $|\Psi_N\rangle$ and violating Selection Rule (\ref{selrul})
is at least of order $D(\vec{\lambda})$, the distance of $\vec{\lambda}$ to the corresponding facet of the polytope.

To gain further insight we explored the most relevant nontrivial setting, the Borland-Dennis setting of $3$ fermions and a $6$-dimensional $1$-particle Hilbert space. Analytically, we confirmed in form of Theorem \ref{lem:xizeta} and Theorem \ref{lem:unstablebound} the stability of Selection Rule (\ref{selrul}) unless the ordering constraint $\lambda_3 \geq \lambda_{4}$ is not approximately saturated. For this we verified that the indicator of stability, $\chi\equiv (1- \|P_{D}\Psi_3\|_{L^2}^2)/D(\vec{\lambda})$, is indeed also bounded from above by $\mathcal{O}(1)$. By sampling the 3-fermion Hilbert space randomly by means of a Monte Carlo method we revealed the tightest possible bounds on the stability which led to an improvement of the bounds in Theorem \ref{lem:unstablebound}.

The potential violation of Selection Rule (\ref{selrul}) is given by a hyperbolic divergence, $1/(\lambda_3-\lambda_{4})$, of the indicator $\chi$. This divergence stated in Theorem \ref{lem:unstablebound} and confirmed analytically requires in general also a modification of Selection Rule (\ref{selrul}) for the case of an additional saturation of an ordering constraint $\lambda_i \geq \lambda_{i+1}$. It depends strongly on the hierarchy of the coefficients $\kappa^{(i)}$ and $\kappa^{(i+1)}$ of the corresponding gPC (cf. Eq.(\ref{gPC})). Based on analytical and numerical findings a modified selection rule has been conjectured (Conjecture \ref{sel2}) for all settings. The random sampling of the $N$-fermion Hilbert space of the higher settings $N=3$, $d=7$ and $N=3$, $d=8$  have confirmed (i) the stability of Selection Rule (\ref{selrul}) in case of well-separated NON, (ii) validity of Conjecture \ref{sel2} and (iii) the divergency of the weight $(1- \|P_{D}\Psi_N\|_{L^2}^2)$ as $1/(\lambda_i-\lambda_{i+1})$.

Due to the particle-hole duality (see e.g.~\cite{RusEquiv}) these results also hold for the settings with $N=4$, $d=7$ and $N=5$, $d=8$. The first non-trivial setting with more than $3$ fermions is given by $\wedge^4[\mathcal{H}_1^{(8)}]$. However, the relative number of quasipinned states for this case is drastically less than for the setting with $N=3$ and $d=8$. Despite this restriction to smaller settings we are convinced that our results (i)-(iii) hold for arbitrary settings $\wedge^N[\mathcal{H}_1^{(d)}]$, as well. Note also that the discussion of degenerate NON is important since (e.g.~orbital) symmetries can lead to degeneracy.

The concrete bounds on the stability of Selection Rule (\ref{selrul}) are essential for applications of the new concept of gPC and quasipinning for concrete fermionic quantum systems. With our results at hand the task of exploring the position of given NON inside of the polytope is not just academic anymore. It allows to draw concrete conclusions on the structure of the corresponding $N$-fermion quantum state $|\Psi_N\rangle$ based on the strength of the quasipinning and the concrete position of the NON in the vicinity of the polytope boundary. In that sense our work provides the basis for a generalized Hartree-Fock method, a variational wave function ansatz using the simplified structure of $|\Psi_N\rangle$ corresponding to pinning.

Due to more involved structure of the polytopes and gPC for larger settings $\wedge^N[\mathcal{H}_1^{(d)}]$ their relevance for physical applications is not obvious. However, the restrictions to the smallest few non-trivial mathematical settings, let's say $\wedge^3[\mathcal{H}_1^{(6)}]$ and $\wedge^3[\mathcal{H}_1^{(7)}]$, does not mean to restrict to three fermion systems with just a $6$- or $7$-dimensional $1$-particle Hilbert space. E.g.~for ground states of few fermion systems (e.g.~electrons in atoms) the conflict between energy minimization and Pauli principle is often so dominant that a small dimensional active space emerges: Some electrons are frozen in lower lying energy shells and all shells with higher energies are never occupied. Indeed, e.g.~for a harmonic system \footnote{C.Schilling, D.Ebler, unpublished} of $N$-fermions with the infinite-dimensional $1$-particle Hilbert space $L^2(\mathbb{R}^3)$ $N-3$ fermions are frozen and the active space is isomorphic to $\wedge^3[\mathcal{H}_1^{(6)}]$. The pinning analysis in this active space has shown that the $N$-fermion ground state is strongly quasipinned. Similar results were found by Klyachko \cite{Kly1} for a Beryllium state where he truncated the pinning analysis to the active space $\wedge^3[\mathcal{H}_1^{(7)}]$.

To conclude, by analytical and numerical means we have shown that Selection Rule (\ref{selrul}) is stable for the case of quasipinning as long as the NON are not quasidegenerate. To deal with the case of (quasi)degenerate NON we conjectured a modified selection rule for pinning. Its uniform stability for the case of quasipinning has been supported analytically and numerically for the settings $N=3$ and $6 \leq d \leq 8$. Since pinning and quasipinning strongly reduces the structure of the $N$-fermion quantum states our work may also provide the basis for a generalized Hartree-Fock method, i.e.~for a variational ansatz using the simplified structure for $|\Psi_N\rangle$ corresponding to pinning.

\section*{Acknowledgments}
We are very grateful to D.\hspace{0.5mm}Gross and A.\hspace{0.5mm}Lopes for several helpful discussions
and editorial comments. We also thank M.\hspace{0.5mm}Christandl, D.\hspace{0.5mm}Jaksch and J.\hspace{0.5mm}Whitfield for helpful discussions and J.C.\hspace{0.5mm}Andresen and J.\hspace{0.5mm}Skowera for their IT support.
We acknowledge financial support from the Swiss National Science Foundation (Grant P2EZP2 152190).
Some of the simulations in this work were performed on the BRUTUS computing cluster at ETH Zurich.

\appendix

\section{Proof of Theorem \ref{stat:ConverseStab}}\label{app:QPlowerbound}
\begin{proof}
We use the spectral decomposition, $\hat{D}_{\Psi_N} = \sum_{\Delta} \Delta\,P_{\Delta}$, where we sum over all eigenvalues $\Delta$ of $\hat{D}_{\Psi_N}$ and denote the projection operator onto the $\Delta$-eigenspace by $P_{\Delta}$. By using the projection operator $P^{(+)}$
onto the positive spectrum of $\hat{D}_{\Psi_N}$ we find
\begin{eqnarray}
D(\vec{\lambda})&=& \langle\Psi_N|\hat{D}_{\Psi_N}|\Psi_N\rangle \nonumber \\
&=& \sum_{\Delta}\, \Delta\,\|P_\Delta \Psi_N\|_{L^2}^2 \nonumber \\
&\leq& \sum_{0<\Delta}\, \Delta\,\|P_\Delta \Psi_N\|_{L^2}^2 \nonumber \\
&\leq & \|\hat{D}_{\Psi_N}\|_{\small{\mbox{op}}} \,\,\|P^{(+)} \Psi_N\|_{L^2}^2 \nonumber \\
&\leq& \|\hat{D}_{\Psi_N}\|_{\small{\mbox{op}}} \,\,(1-\|P_D \Psi_N\|_{L^2}^2)\,.
\end{eqnarray}
\end{proof}

\section{Proof of Theorem \ref{lem:xizeta}}\label{app:xizeta}
\begin{proof}
Consider $|\Psi_3\rangle\in \wedge^3[\mathcal{H}_1^{(6)}]$ expanded according to Eq.~(\ref{8SD}).
The gPC (\ref{d=6b}) takes the form
\begin{equation}\label{beta0}
D^{(3,6)}(\vec{\lambda}) = -|\beta|^2 + |\gamma|^2 +|\delta|^2 +2|\xi|^2 +|\zeta|^2 \geq 0,
\end{equation}
and thus we find
\begin{equation}\label{beta}
|\beta|^2 = -D^{(3,6)}(\vec{\lambda}) + |\gamma|^2 +|\delta|^2 +2|\xi|^2 +|\zeta|^2 \,.
\end{equation}
Notice again the strength of the geometric picture. By recalling the normal vector $(-1,1,1)$ for the (red) plane/facet $F_{D^{(3,6)}}$ in
Fig.~\ref{fig:polytope}, Eq.~(\ref{beta0}) follows immediately geometrically.
Moreover, in Fig.~\ref{fig:polytope} we geometrically observe that $\Delta\lambda\equiv \lambda_3-\lambda_4\geq 0$ becomes
\begin{equation}\label{delta34}
\Delta \lambda = |\alpha|^2+ |\gamma|^2+|\delta|^2+|\xi|^2-|\beta|^2-|\mu|^2-|\nu|^2-|\zeta|^2\,,
\end{equation}
which yields
\begin{equation}\label{alpha}
|\alpha|^2 = \Delta \lambda -D^{(3,6)}(\vec{\lambda}) + |\mu|^2+|\nu|^2  + |\xi|^2+2|\zeta|^2\,.
\end{equation}
The diagonality of $\rho_1$ w.r.t.~the NO $|k\rangle$ in particular means that $\langle3|\rho_1|4\rangle =0$, which leads to
\begin{equation}\label{1RDO34}
-\alpha \beta^\ast = \gamma \nu^\ast + \delta \mu^\ast + \xi \zeta^\ast\,.
\end{equation}
The idea is now to make use of the fact that for quasipinning, $D^{(3,6)}(\vec{\lambda})\approx 0$, the $\alpha$ and $\beta$ point in Fig.~\ref{fig:polytope} carry the main weight according to Eqs.~(\ref{beta}), (\ref{alpha}). Indeed, geometrically quasipinning means that $\vec{v}$
is close to the (red) plane and therefore in particular on the same side of the (green) plane as the $\alpha$-point (or maximally $D^{(3,6)}(\vec{\lambda})$-far away on the other side). This means mathematically that $|\beta|^2 \approx |\gamma|^2+|\delta|^2+ 2 |\xi|^2 + |\zeta|^2$ and $|\alpha|^2 \gtrsim |\mu|^2+|\nu|^2+  |\xi|^2 + 2 |\zeta|^2$, which together with the normalization implies $|\alpha|^2+|\beta|^2 \gtrsim \frac{1}{2}$.
From that viewpoint Eq.~(\ref{1RDO34}) can imply a quite restrictive condition.
By defining $\vec{v}\equiv(\nu,\mu,\zeta)$ and $\vec{w}\equiv(\gamma,\delta,\xi)$ and using the Cauchy-Schwartz inequality
we find for Eq.~(\ref{1RDO34})
\begin{equation}
|\alpha|^2 |\beta|^2 = |\langle \vec{v},\vec{w}\rangle|^2\leq |\vec{v}|^2 |\vec{w}|^2\,.
\end{equation}
This together with Eqs.~(\ref{beta}) and (\ref{alpha}) yields
\begin{widetext}
\begin{equation}
\left(|\vec{v}|^2+|\xi|^2 +|\zeta|^2+\Delta \lambda-D^{(3,6)}(\vec{\lambda})\right)\left(|\vec{w}|^2+|\xi|^2 +|\zeta|^2-D^{(3,6)}(\vec{\lambda})\right)\leq |\vec{v}|^2 |\vec{w}|^2\,.
\end{equation}
This leads to
\begin{eqnarray}\label{ineq1}
0 &\geq & \left(|\xi|^2 +|\zeta|^2-D^{(3,6)}(\vec{\lambda})\right)\left(|\vec{v}|^2+|\vec{w}|^2\right) + \left(|\xi|^2 +|\zeta|^2-D^{(3,6)}(\vec{\lambda}\right)^2 + \Delta \lambda \left(|\vec{w}|^2+|\xi|^2 +|\zeta|^2-D^{(3,6)}(\vec{\lambda})\right)\,.
\end{eqnarray}
\end{widetext}
In particular, since $\left(|\vec{w}|^2+|\xi|^2 +|\zeta|^2-D^{(3,6)}(\vec{\lambda})\right) = |\beta|^2, \Delta \lambda\geq 0$ this requires
\begin{equation}
|\xi|^2 + |\zeta|^2 \leq D^{(3,6)}(\vec{\lambda})\,.
\end{equation}
\end{proof}

\section{Proof of Theorem \ref{lem:unstablebound}}\label{app:unstablebound}
\begin{proof}
The proof will be based on the estimates we have already found in the proof of Theorem \ref{lem:xizeta}. Due to Eq.~(\ref{beta}) we first estimate the term
$|\gamma|^2+|\delta|^2$. Recasting Eq.~(\ref{ineq1}) yields
\begin{widetext}
\begin{eqnarray}
\Delta \lambda \, |\vec{w}|^2 &\leq & \left(D^{(3,6)}(\vec{\lambda})-|\xi|^2 -|\zeta|^2\right) \left(|\vec{v}|^2+|\vec{w}|^2 -D^{(3,6)}(\vec{\lambda})+|\xi|^2 +|\zeta|^2\right) + \Delta \lambda \left(D^{(3,6)}(\vec{\lambda})-|\xi|^2 -|\zeta|^2\right)
\end{eqnarray}
Then, we estimate  (recall that $|\vec{w}|^2 = |\gamma|^2+|\delta|^2+|\xi|^2$)
\begin{eqnarray}\label{ineq2}
\Delta \lambda \,(|\gamma|^2+|\delta|^2) &\leq & D^{(3,6)}(\vec{\lambda}) \left(|\vec{v}|^2+|\vec{w}|^2 -D^{(3,6)}(\vec{\lambda})+|\xi|^2 +|\zeta|^2\right) + \Delta \lambda \,D^{(3,6)}(\vec{\lambda}) \,.
\end{eqnarray}
\end{widetext}
Moreover, by using Eqs.~(\ref{beta0}), (\ref{delta34}) and  $\vec{v}\equiv(\nu,\mu,\zeta)$, $\vec{w}\equiv(\gamma,\delta,\xi)$
we find
\begin{eqnarray}
|\vec{v}|^2+|\vec{w}|^2 -D^{(3,6)}(\vec{\lambda})+|\xi|^2 +|\zeta|^2 &=& |\beta|^2 +|\mu|^2+|\nu|^2 +|\zeta|^2 \nonumber \\
&=& \lambda_4 \,\,\,\leq\,\,\, \frac{1}{2}\,.
\end{eqnarray}
Finally, with Eqs.~(\ref{beta}), (\ref{ineq2}), Theorem \ref{lem:xizeta} and $\Delta \lambda \equiv \lambda_3-\lambda_4$ this leads to
\begin{eqnarray}
|\beta|^2 +|\gamma|^2 + |\delta|^2  &\leq& 2 (|\gamma|^2 + |\delta|^2) -D^{(3,6)}(\vec{\lambda}) +2 |\xi|^2 +|\zeta|^2 \nonumber \\
&\leq& 2 (|\gamma|^2 + |\delta|^2) +D^{(3,6)}(\vec{\lambda})  \nonumber \\
&\leq& \frac{D^{(3,6)}(\vec{\lambda})}{\lambda_3-\lambda_4}+3 D^{(3,6)}(\vec{\lambda})\,.
\end{eqnarray}
\end{proof}

\bibliography{bibliography}

\end{document}